%% file: rangeni.tex
\input{rangeni.def}
\documentstyle[12pt]{article}
\title{\sf Harmonic analysis of random number generators} %l.4
\author{{\sc Oliver Schnetz}\thanks{Institut f{\"u}r theoretische %l.5 2{
Physik III, Staudtstra{\ss}e 7, 91058 Erlangen, Germany,\newline %l.6 2{
e-mail: schnetz@pest.physik.uni-erlangen.de\newline %l.7 2{
Supported in parts by the DFG Graduiertenkolleg 'Starke Wechselwirkung' %l.8 2{
and the BMBF.\newline %l.9 2{
FAU-TP3-96/14}}\date{September 26, 1996}%l.10
\begin{document}\renewcommand{\arraystretch}{1.2} %l.11
\maketitle %l.12
\begin{abstract} %l.13 úBAt
The spectral test of random number generators (R.R. Coveyou and R.D. %l.14 úBAt
McPherson, 1967) is generalized. The sequence of random numbers %l.15 úBAt
is analyzed explicitly, not just via their $n$-tupel distributions. %l.16 úBAt
We find that the mixed multiplicative generator with power of two %l.17 úBAt
modulus does not pass the extended test with an ideal result. Best %l.18 úBAt
qualities has a new generator with the recursion formula $ X_{k%l.19 ó,ò...
+1}=aX_{k}+c{\rm\hspace{.38ex}int}(k/2){\rm\hspace{.38ex}mod\hspace{.38ex}}2^d%l.20 $...
$. We discuss the choice of the parameters $a$, $c$ for very large %l.21 úBAt
moduli $ 2^d$ and present an implementation of the suggested generator %l.22 úBAt
with $ d=256$, $ a=2^{128}+2^{64}+2^{32}+62181$, $ c=(2^{160}+1%l.23 $...
)\cdot 11463$.%l.24 úBAt
\end{abstract} %l.25
\tableofcontents %l.26
 %l.27
\section{Introduction} %l.28
The spectral test was proposed by R.R. Coveyou and R.D. McPherson %l.29
in 1967 \cite{Cov}. The advantage of this test is to present an %l.30
algebraic criterion for the quality of the generator. For the mixed %l.31
multiplicative generator $ X_{k+1}=aX_{k}+c{\rm\hspace{.38ex}mod\hspace{.38ex}}M%l.32 $
$ %l.33
\begin{equation}%l.34 ú$
\label{03}{\rm\hspace{.38ex}min\hspace{.38ex}}\{|{\bf s}|=\sqrt{s%l.35 ...
_{1}^{2}+{\ldots}+s_{n}^{2}{}} {\rm  \ with \ }s_{a}=s_{1}+as_{%l.36 ó,ò...
2}+{\ldots}+a^{n-1}s_{n}=0{\rm\hspace{.38ex}mod\hspace{.38ex}}M%l.37 ú$
\}%l.38 ú$
\end{equation} %l.39
should be as large as possible \cite{Cov,Knu}. The criterion is that %l.40
simple since the $n$-tupels of random numbers form an $n$-dimensional %l.41
lattice (cf.\ e.g.\ Fig.\ 2). A good generator has uniformly distributed %l.42
$n$-tupels which refers to an almost cubic lattice \cite{M,B,Nie}.%l.43

The lattice is a consequence of the (affine) linear dependence of %l.45
$X$$_{k+1}$ on $X$$_{k}$. From the figures on the left (type I) %l.46
we see that the relation between $k$ and $X$$_{k}$ is much more %l.47
complicated. This is however one of the most fundamental aspects %l.48
of randomness. In order to judge whether a sequence $X$$_{k}$ takes %l.49
random values one would first plot the sequence itself and then %l.50
maybe $X$$_{k+1}$ over $X$$_{k}$.%l.51

Of course, the correlation between $k$ and $X$$_{k}$ is not independent %l.53
from the distribution of pairs $ (X_{k},X_{k+1})$. E.g., a poor %l.54
'random' sequence $ X_{k}=ak$ lying on a line with gradient $a$ %l.55
leads to pairs $ (X_{k},X_{k+1}=X_{k}+a)$ lying on a line with gradient %l.56
1 shifted by $a$ off the origin. This makes it reasonable to judge %l.57
randomness by only looking at the $n$-tupel distributions. However, %l.58
random number generators which have identical valuation by the spectral %l.59
test may still look quite different. The generators $ X_{k+1}=41X%l.60 $
_{k}+3{\rm\hspace{.38ex}mod\hspace{.38ex}}1024$ (Fig.\ 3) and $ X%l.61 $
_{k+1}=41X_{k}+1{\rm\hspace{.38ex}mod\hspace{.38ex}}1024$ (Fig.\ %l.62
4), e.g., differ only by the additive constant which does not enter %l.63
Eq.\ (\ref{03}). The lattices of pair distributions (type II in %l.64
the figures) are similar whereas the plots of $X$$_{k}$ over $k$ %l.65
show different behavior. The spectral test not even makes a difference %l.66
between a prime number and a power of two modulus (cf.\ Fig.\ 1 %l.67
vs.\ Fig.\ 2).%l.68

Therefore it is desirable to include the analysis of the correlation %l.70
between $k$ and $X$$_{k}$ into the valuation of the test. In fact %l.71
it is possible to analyze the accumulation of random numbers along %l.72
certain lines (which is often seen in the figures) by Fourier transformation. %l.73
More generally we extend the spectral test by analyzing the correlation %l.74
between $k$ and the $n$-tupel $ (X_{k},X_{k+1},{\ldots},X_{k+n-%l.75 ó,ò...
1})$. The generators mentioned above (Figs.\ 3, 4) acquire different %l.76
valuations. Fig.\ 4 is preferred since the random numbers spread %l.77
more uniformly in Fig.\ 4I than in Fig.\ 3I (cf.\ Sec.\ \ref{mm} %l.78
1).%l.79

We will find that the commonly used mixed multiplicative generator %l.81
always shows correlations along certain lines if the modulus is %l.82
not a prime number. We will present an improved generator which %l.83
is almost free from these correlations (Fig.\ 7). It has the recursion %l.84
formula %l.85
\begin{equation}%l.86 ú$
\label{02}X_{k+1}=aX_{k}+c{\rm\hspace{.38ex}int}\left( k/2%l.87 ú$
\right) {\rm\hspace{.38ex}mod\hspace{.38ex}}2^d%l.88 ú$
\end{equation} %l.89
with the parameters %l.90
\begin{eqnarray}%l.91 ú$$
d=2^{k}d_{0}&,&a=2^{2^{k-1}d_{0}}+2^{2^{k-2}d_{0}}+{\ldots}+2^{2d%l.92 { ú$$
_{0}}+\left( 3~580~621~541{\rm\hspace{.38ex}mod\hspace{.38ex}}2^%l.93 { (,)...
{d_{0}}\right) ,\nonumber \\ %l.94 ú$$
\label{01}&&c=\left( 2^{{\rm int}\left( 2^{k+1}{} /3\right) d_{0%l.95 ó,ò...
}}+1\right) \left( 3~370~134~727{\rm\hspace{.38ex}mod\hspace{.38ex}}2^%l.96 { (,)...
{d_{0}}\right) .%l.97 ú$$
\end{eqnarray} %l.98
In particular, the case $ d_{0}=16$, $ k=4$ is discussed in Ex.\ %l.99
7.1.%l.100

This generator is supposed to be a good choice with respect to the %l.102
following three criteria.%l.103

Firstly, the sequence of numbers provided by the generator should %l.105
behave as close to a true random sequence as possible.%l.106

Secondly, the calculation of random numbers should be as fast as %l.108
possible. The generator given in Eqs.\ (\ref{02}), (\ref{01}) is %l.109
explicitly constructed to have best performance. It is important %l.110
to note that this is not independent from the first criterion. It %l.111
is possible to produce better random numbers the more effort one %l.112
spends in calculating the numbers. Figs.\ 2 and 6 show how a simple %l.113
doubling of the digits of the modulus improves the randomness of %l.114
the generator. In general we can produce arbitrarily good random %l.115
numbers with e.g.\ $ d_{0}=32$ and large $k$ in Eq.\ (\ref{01}).%l.116

Thirdly, the properties of the random numbers should be known as %l.118
detailed as possible. It is not sufficient to use a messy, opaque %l.119
formula. It has often been seen that this leads to numbers which %l.120
are far from being random \cite{Knu}. As long as one is not familiar %l.121
with the qualities of the generator one can never rely on the results %l.122
gained with it. The full evaluation of the generalized spectral %l.123
test is supposed to provide a profound knowledge of the generator.%l.124

\vspace{1ex}%l.126
\noindent{}We start with the development of the generalized spectral %l.127
test in the next section. In Sec.\ \ref{generators} we apply the %l.128
test to a series of commonly used and some new generators. Finally %l.129
we discuss the choice of parameters in Sec.\ \ref{param}.%l.130

\section{The generalized spectral test} %l.132
\subsection{Review of the spectral test} %l.133
We start with a short review of the spectral test \cite{Cov,Knu} %l.134
in which we try to stress its geometrical meaning. The idea is to %l.135
plot all $n$-tupels of successive random numbers in an $n$-dimensional %l.136
diagram. This is done, e.g.\ for $ n=2$ in the figures of type II.%l.137

Mathematically a figure is presented as a function $g$ which is 1 %l.139
at every dot and 0 elsewhere. If $ \NX$ is the period of the generator %l.140
$X$, that is the smallest number with $ X_{k+\NX }=X_{k}%l.141 $
\hspace*{1ex}\forall k$, then %l.142
\begin{equation}%l.143 ú$
g\left( x_{1},{\ldots},x_{n}\right) =\sum _{k=1}^\NX \delta _{x_{%l.144 ó,ò...
1},X_{k}}\cdots \delta _{x_{n},X_{k+n-1}}\equiv \sum _{k\in %l.145 { ú$
{\ErgoBbb Z}_\NX }\delta _{{\bf x},{\bf X}_{k}}\hspace{.6ex},%l.146 ú$
\end{equation} %l.147
where $ \delta _{a,b}=1$ if $ a=b$ and $ \delta _{a,b}=0$ if $ a%l.148 $
\neq b$ (for later convenience we also write the Kronecker $%l.149
\delta $ as $ \delta _{a=b}$). Moreover we have introduced the notation %l.150
\begin{equation}%l.151 ú$
{\bf X}_{k}=\left( X_{k},X_{k+1},{\ldots},X_{k+n-1}\right) %l.152 ú$
\hspace{.6ex},\hspace{2ex}{\bf x}=\left( x_{1},x_{2},{\ldots},x_{%l.153 ó,ò...
n}\right) \hspace{.6ex},\hspace{2ex}{\ErgoBbb Z}_\NX =%l.154 ú$
{\ErgoBbb Z}/\NX{\ErgoBbb Z}\hspace{.6ex}.%l.155 ú$
\end{equation}%l.156

We want to check whether the dots accumulate along certain hyper-planes %l.158
(see e.g.\ the lines in Fig.\ 3II). To this end we select a hyper-plane %l.159
and project all the dots onto a line perpendicular to it. If points %l.160
accumulate along the plane many dots will lie on top of each other, %l.161
otherwise the dots are spread uniformly over the line.%l.162

The hyper-plane $H$ is determined by its Hesse normal form %l.164
\begin{equation}%l.165 ú$
H=\{{\bf x}:s_{1}x_{1}+s_{2}x_{2}+{\ldots}+s_{n}x_{n}\equiv %l.166 ú$
{\bf s}\cdot {\bf x}=0\}\hspace{.6ex},\hspace{2ex} |{\bf s}|%l.167 ú$
\equiv \sqrt{{\bf s}\cdot {\bf s}} \equiv \sqrt{s_{1}^{2}+%l.168 ...
{\ldots}+s_{n}^{2}{}} \neq 0\hspace{.6ex}.%l.169 ú$
\end{equation} %l.170
The line is stretched by the factor $ |{\bf s}|$. The position of %l.171
a point $ {\bf X}_{k}$ on the line perpendicular to the plane is %l.172
given by the number $ {\bf s}\cdot {\bf X}_{k}$.%l.173

Next we wind the line up to a circle so that the modulus $M$ as point %l.175
on the line lies on top of the 0. For a suitable choice of $ %l.176 $
{\bf s}$, namely $ {\bf s}=(-1,25)$ all points in Fig.\ 3II lie %l.177
now on the point represented by the number 25.%l.178

The points are realized as complex phases on the unit circle. We %l.180
obtain the assignment %l.181
\begin{equation}%l.182 ú$
{\bf X}_{k}\mapsto \exp\left( \frac{2\pi i}{M} {\bf s}\cdot %l.183 (,)...
{\bf X}_{k}\right) \hspace{.6ex}.%l.184 ú$
\end{equation} %l.185
Finally we draw arrows from the center of the circle to all the dots %l.186
and add them. The length of the resulting vector describes how the %l.187
dots are balanced on the circle. If the dots spread uniformly the %l.188
arrows cancel each other and the resulting vector is small. If, %l.189
on the other hand, all dots lie on top of each other the length %l.190
of the arrows sums up to $ \NX$.%l.191

If we restrict ourselves to integer $s$$_{1}$, {\dots}, $s$$_{n}$ %l.193
(accumulation of random numbers always occur along hyper-planes %l.194
given by integer $s$$_{i}$) the resulting vector is given by the %l.195
Fourier transform of $g$, %l.196
\begin{equation}%l.197 ú$
\hat{g}\left( {\bf s}\right) =\frac{1}{\sqrt{\NX}} \sum _{%l.198 { ú$
{\bf x}\in {\ErgoBbb Z}^{n}_\NX }g\left( {\bf x}\right) \exp%l.199 (,)...
\left( \frac{2\pi i}{M} {\bf s}\cdot {\bf x}\right) =\frac{1}{%l.200 ...
\sqrt{\NX}} \sum _{k\in {\ErgoBbb Z}_\NX }\exp\left( \frac{2%l.201 ó,ò...
\pi i}{M} {\bf s}\cdot {\bf X}_{k}\right) \hspace{.6ex},%l.202 ú$
\end{equation} %l.203
where we have introduced the normalization factor $ \NX^{-1/2}$. %l.204
The information about accumulations along the hyper-plane is contained %l.205
in $ |\hat{g}|^{2}$, the phase of $ \hat{g}$ is irrelevant.%l.206

\vspace{1ex}%l.208
\noindent{}We remember that for the mixed multiplicative generator %l.209
the $n$-tupels form a lattice (which is displaced off the origin). %l.210
So $ |\hat{g}|^{2}({\bf s})$ will assume the maximum value $ \NX%l.211 $
$ if $ {\bf s}$ lies in the dual lattice, $ {\bf s}\cdot %l.212 $
{\bf X}_{k}=C+\ell M$, $ C,\ell \in {\ErgoBbb Z}$, otherwise $ %l.213 $
|\hat{g}|^{2}({\bf s})$ is zero. Since $ X_{k}=c(a^{k}-1)/(a-1){\rm\hspace%l.214 `...
{.38ex}mod\hspace{.38ex}}M$ this means, if $ {\rm\hspace%l.215 `...
{.38ex}gcd}(c,M)=1$ and $X$ has full period, that $ \forall k:%l.216 $
\hspace*{1ex}(a^{k}-1)s_{a}=0{\rm\hspace{.38ex}mod\hspace{.38ex}}M%l.217 $
{\rm\hspace{.38ex}gcd}(a-1,M)$ from which Eq.\ (\ref{03}) follows %l.218
(cf.\ Sec.\ \ref{mm}).%l.219

\subsection{Generalization of the spectral test\label{gentest}} %l.221
We generalize the spectral test by caring for the sequence in which %l.222
the $n$-tupels are generated. The index $k$ is added to the $n$-tupel %l.223
$ {\bf X}_{k}$ as zeroth component and we define $g$ as %l.224
\begin{equation}%l.225 ú$
g\left( x_{0},{\bf x}\right) =\sum _{k\in {\ErgoBbb Z}_\NX }%l.226 ú$
\delta _{x_{0},k}\delta _{{\bf x},{\bf X}_{k}}=\delta _{{\bf x}%l.227 { ú$
,{\bf X}_{x_{0}}}\hspace{.6ex}.%l.228 ú$
\end{equation} %l.229
The geometrical interpretation remains untouched but now we consider %l.230
also the figures of type I. The Fourier transform of $g$ is given %l.231
by %l.232
\begin{equation}%l.233 ú$
\label{1}\hat{g}\left( s_{0},{\bf s}\right) =\frac{1}{\sqrt{\NX}%l.234 ...
} \sum _{k\in {\ErgoBbb Z}_\NX }\exp\left( \frac{2\pi i}{\NX} s_{%l.235 ó,ò...
0}k+\frac{2\pi i}{M} {\bf s}\cdot {\bf X}_{k}\right) %l.236 ú$
\hspace{.6ex}.%l.237 ú$
\end{equation} %l.238
The sum over $k$ is hard to evaluate since in the exponential $k$ %l.239
is combined with $X$$_{k}$. However in fact we are interested in %l.240
$ |\hat{g}|^{2}$ and find %l.241
\begin{eqnarray}%l.242 ú$$
|\hat{g}|^{2}\left( s_{0},{\bf s}\right) &=&\frac{1}{\NX} \sum _%l.243 { ú$$
{k,k'\in {\ErgoBbb Z}_\NX }\exp\left( \frac{2\pi i}{\NX} s_{0}%l.244 (,)...
\left( k'-k\right) +\frac{2\pi i}{M} {\bf s}\cdot \left( %l.245 (,)...
{\bf X}_{k'}-{\bf X}_{k}\right) \right) \nonumber \\ %l.246 ú$$
\label{15}&=&\frac{1}{\NX} \sum _{\Delta k\in {\ErgoBbb Z}_\NX %l.247 ú$$
}\exp\left( \frac{2\pi i}{\NX} s_{0}\Delta k\right) \sum _{k%l.248 { ú$$
\in {\ErgoBbb Z}_\NX }\exp\left( \frac{2\pi i}{M} {\bf s}\cdot %l.249 (,)...
\left( {\bf X}_{k+\Delta k}-{\bf X}_{k}\right) \right) %l.250 ú$$
\hspace{.6ex}.%l.251 ú$$
\end{eqnarray} %l.252
The sum over $k$ has no linear $k$-dependence, only differences of %l.253
random numbers occur. Like in the standard spectral test in many %l.254
cases the sum over $k$ can be evaluated. The result is often simple %l.255
enough to be able to evaluate the sum over $ \Delta k$ also.%l.256

Note that the standard spectral test corresponds to $ s_{0}=0$. We %l.258
give some simple results on $ |\hat{g}|^{2}$ in the following lemma.%l.259

\pagebreak[3]%l.261

\noindent {\bf Lemma} 2.1. %l.263
\begin{eqnarray}%l.264 ú$$
\label{20}|\hat{g}|^{2}[X_{k+c_{2}}+c_{3}]\left( s_{0},{\bf s}%l.265 ú$$
\right) &=&|\hat{g}|^{2}[X_{k}]\left( s_{0},{\bf s}\right) %l.266 ú$$
\hspace{.6ex},\\ %l.267 ú$$
\label{29}|\hat{g}|^{2}\left( s_{0},{\bf 0}\right) &=&\NX %l.268 ú$$
\delta _{s_{0}=0{\rm\hspace{.38ex}mod}\NX }\hspace{.6ex},\\ %l.269 ú$$
\label{30}\sum _{s_{0}\in {\ErgoBbb Z}_\NX }|\hat{g}|^{2}%l.270 (,)...
\left( s_{0},{\bf s}\right) &=&\NX\\ %l.271 ú$$
\sum _{{\bf s}\in {\ErgoBbb Z}_M^{n}}|\hat{g}|^{2}\left( s_{0},%l.272 (,)...
{\bf s}\right) &=&M^{n}\hspace*{1cm}\hbox{\hspace{.38ex}if }%l.273 ú$$
{\bf X}_{k}={\bf X}_{k'}\Rightarrow k=k'{\rm\hspace{.38ex}mod}\NX %l.274 ú$$
\hspace{.6ex}.%l.275 ú$$
\end{eqnarray} %l.276
One may also be interested in correlations between non-successive %l.277
random numbers like $X$$_{k}$ and $X$$_{k+2}$. In general it is %l.278
possible to study $n$-tupels $ {\bf X}_{k+\tau }\equiv (X_{k+%l.279 { $
\tau _{1}},{\ldots},X_{k+\tau _{n}})$. This amounts to replacing %l.280
$ {\bf X}_{k}$ by $ {\bf X}_{k+\tau }$ and $s$$_{a}$ by $ s_{a,%l.281 { $
\tau }\equiv s_{1}a^{\tau _{1}}+{\ldots}+s_{n}a^{\tau _{n}}$ in %l.282
our results.%l.283

\subsection{Valuation with the generalized spectral test} %l.285
Now we have to clarify how the calculation of $ |\hat{g}|^{2}$ leads %l.286
to a valuation of the generator.%l.287

We can not expect that $ |\hat{g}|^{2}$ vanishes identically outside %l.289
the origin since in this case $g$ would be constant. Eq.\ (\ref{30}) %l.290
shows that the mean value of $ |\hat{g}|^{2}$ is 1.%l.291

What would we expect for a sum of truly random phases? Real and imaginary %l.293
part of a random arrow with length 1 have equal variance $ 1/2$. %l.294
For large $ \NX$ the sum of arrows is therefore normally distributed %l.295
with density $ 1/\pi \NX\cdot \exp(-(x^{2}+y^{2})/\NX )dxdy=\exp%l.296 $
(-r^{2}{} /\NX )dr^{2}{} /\NX$. Thus $ z=|\hat{g}|^{2}$ has the %l.297
density $ \exp(-z)$ for a true random sequence and the expected %l.298
value for $ |\hat{g}|^{2}$ is 1.%l.299

This means that values of $ |\hat{g}|^{2}\le 1$ can be accepted. %l.301
It is clear that for a given $ (s_{0},{\bf s})\neq (0,{\bf 0})$ %l.302
the correlations are worse the higher $ |\hat{g}|^{2}(s_{0},%l.303 $
{\bf s})>1$ is. But what does the location of an $ (s_{0},%l.304 $
{\bf s})$ with $ |\hat{g}|^{2}(s_{0},{\bf s})>1$ mean for the generator?%l.305

We remember that $ (s_{0},{\bf s})$ may be seen as normal vector %l.307
on the hyper-plane along which the accumulations occur. If e.g.\ %l.308
$ n=1$ and $ (s_{0},s_{1})=(1,1)$ the corresponding 1-plane has %l.309
the equation $ x_{0}+x_{1}=0$ (cf.\ e.g.\ Figs.\ 2I, 3I). If the %l.310
$k$-axis and the $X$$_{k}$-axis are normalized to length 1 this %l.311
line has length $ \sqrt{2}$. With the normal vector (3,1) (cf.\ %l.312
Fig.\ 4I) one obtains the equation $ 3x_{0}+x_{1}=0{\rm\hspace%l.313 `...
{.38ex}mod\hspace{.38ex}}1$ which intersects the unit cube three %l.314
times and therefore has the length $ \sqrt{3^{2}+1} =\sqrt{10}$. %l.315
Accumulations along this longer line are less important than along %l.316
the short line. In the extreme case where the line fills the whole %l.317
unit cube by intersecting it very often, accumulations can hardly %l.318
be recognized. Note that in this sense the normal vectors $ (s_{%l.319 ó,ò...
0},s_{1})$ and $ (2s_{0},2s_{1})$ do not determine the same line. %l.320
The latter one contains e.g.\ the points $ (1/2,0)$, $ (0,1/2)$, %l.321
$ (1,1/2)$, $ (1/2,1)$. It has twice the length of the former one %l.322
and too large a $ |\hat{g}|^{2}$ has half the effect.%l.323

We generalize these considerations to $ n>1$ by taking the area of %l.325
the $n$-dimensional hyper-plane with normal vector $ (s_{0},%l.326 $
{\bf s})$ as measure for the importance of the accumulations detected. %l.327
The area is given by $ |(s_{0},{\bf s})|=(s_{0}^{2}+{\bf s}^{2}%l.328 $
)^{1/2}$, the Euclidean length of the normal vector.%l.329

We can relate both mechanisms by defining the quality parameter %l.331
\begin{equation}%l.332 ú$
\label{55}Q_{n}\left( s_{0},{\bf s}\right) \equiv \frac{|\left( s%l.333 (,)...
_{0},{\bf s}\right) |}{|\hat{g}\left( s_{0},{\bf s}\right) |^{2%l.334 ó,ò...
}{}} \hspace{.6ex},\hspace{2ex}Q_{n}\equiv {\rm\hspace{.38ex}max\hspace{.38ex}}_%l.335 { ú$
{\left( s_{0},{\bf s}\right) \in {\ErgoBbb Z}_\NX \times %l.336 { ú$
{\ErgoBbb Z}_M^{n}\backslash \{0,{\bf 0}\}}Q_{n}\left( s_{0},%l.337 (,)...
{\bf s}\right) \hspace{.6ex}.%l.338 ú$
\end{equation} %l.339
Good generators have $ Q_{1}\approx 1$. It is hard to achieve $ Q%l.340 $
_{n}\approx 1$ for $ n>1$ (see however Sec.\ \ref{multrec}). More %l.341
realistic is $ Q_{n}\approx M^{1/n-1}$ (cf.\ Sec.\ \ref{param}) %l.342
which means that the distribution of $n$-tupels deteriorates for %l.343
higher $n$. In general small $n$ are more important than large $n$. %l.344
Apart from the value of $Q$$_{n}$ also the number of sites $ (s_{%l.345 ó,ò...
0},{\bf s})$ at which $ Q_{n}(s_{0},{\bf s})=Q_{n}$ is relevant %l.346
(cf.\ Sec.\ \ref{mm} 1.).%l.347

Let us try to find an interpretation for $Q$$_{n}$. Assume the generator %l.349
produces only multiples of $ t|M$. Then $ |\hat{g}|^{2}(0,s_{1}%l.350 $
=M/t,0,{\ldots},0)=\NX$, thus $ Q_{n}(0,M/t,0,{\ldots},0)=M/t\NX%l.351 $
$, and $ \NX Q_{n}$ determines the number of non-trivial digits. %l.352
In general $ \NX Q_{n}$ may be larger than $M$ and therefore we %l.353
say that $ \tilde{M}_{n}\equiv {\rm\hspace{.38ex}max}(\NX Q_{n}%l.354 $
,M)$ determines the number of digits we can rely on. Analogously %l.355
$ \tilde{N}_{n}\equiv {\rm\hspace{.38ex}max}(\NX Q_{n},\NX)$ gives %l.356
the quantity of random numbers for which the $n$-tupel distributions %l.357
are reasonably random. Specifically $ \tilde{M}_{n}(s_{0},%l.358 $
{\bf s})={\rm\hspace{.38ex}max}(\NX Q_{n}(s_{0},{\bf s}),M)$ determines %l.359
the digits and $ \tilde{N}_{n}(s_{0},{\bf s})={\rm\hspace%l.360 `...
{.38ex}max}(\NX Q_{n}(s_{0},{\bf s}),\NX )$ the quantity of random %l.361
numbers not affected by accumulations perpendicular to $ (s_{0}%l.362 $
,{\bf s})$ (cf.\ \cite[,p. 90]{Knu}).%l.363

Note however that these are only crude statements. If, e.g., the %l.365
'period' of the generator is enlarged by simply repeating it then %l.366
$ \NX Q_{n}(s_{0},{\bf s})$ remains unaffected only if $ s_{0}=0%l.367 $
$. Moreover a high $ |\hat{g}|^{2}(s_{0},{\bf s})$ may be harmful %l.368
even if $ |(s_{0},{\bf s})|$ is large.%l.369

Note that $Q$$_{n}$ is a relative quality parameter. Although $Q$%l.371
$_{n}$ usually does not increase with larger modulus (for $ n>1%l.372 $
$ is actually decreases) the quality of the generator gets better %l.373
since $ \NX$ grows (cf.\ Fig.\ 2 vs.\ Fig.\ 6).%l.374

\section{Generators\label{generators}} %l.376
Here we restrict ourselves to the analysis of the most important %l.377
generators. More examples are found in \cite{Schnetz}. %l.378
\subsection{$ X_{0}=1$, $ X_{k+1}=aX_{k}{\rm\hspace{.38ex}mod\hspace{.38ex}}P%l.379 $
$\label{mul}} %l.380
Let $P$ be a prime number and $a$ a primitive element of $ %l.381 $
{\ErgoBbb Z}_P^\times $, the multiplicative group of $ %l.382 $
{\ErgoBbb Z}_P$ (Fig.\ 1).%l.383

We start the analysis of this multiplicative generator with Eq.\ %l.385
(\ref{15}). We find $ {\bf s}\cdot ({\bf X}_{k+\Delta k}-%l.386 $
{\bf X}_{k})=s_{a}a^{k}(a^{\Delta k}-1)=s_{a}\tilde{k}(a^{%l.387 { $
\Delta k}-1){\rm\hspace{.38ex}mod\hspace{.38ex}}P$ for some $ 0%l.388 $
\neq \tilde{k}\in {\ErgoBbb Z}_P$. If $k$ runs through the $ P-1%l.389 $
$ values of $ {\ErgoBbb Z}_P^\times $ then $ \tilde{k}$ sweeps out %l.390
the whole $ {\ErgoBbb Z}_P\backslash \{0\}$. Assume $ s_{a}%l.391 $
\neq 0$ (the case $ s_{a}=0$ is trivial), then the sum over $ %l.392 $
\tilde{k}$ can be evaluated yielding $ P\delta _{\Delta k=0}-1$. %l.393
Finally the sum over $ \Delta k$ gives together with the normalization %l.394
$ |\hat{g}|^{2}=P/(P-1)-\delta _{s_{0}=0}$. %l.395
\begin{equation}%l.396 ú$
\label{36}%l.397 úBA...
\begin{array}{c|cc}|\hat{g}|^{2}\left( s_{0},{\bf s}\right) &s_{%l.398 ó,ò...
0}=0&s_{0}\neq 0\\\hline  %l.399 úBA...
s_{a}=0&P-1&1/\left( P-1\right) \\ %l.400 úBA...
s_{a}\neq 0&0&P/\left( P-1\right) %l.401 úBA...
\end{array}%l.402 ú$
\end{equation} %l.403
We find that $ Q_{1}=Q_{1}(1,1)=\sqrt{2} (P-1)/P$ is independent %l.404
of $a$ and even greater than 1.\ For $ n\ge 2$ only the case $ %l.405 $
(s_{0},s_{a})=(0,0)$ contributes to $Q$$_{n}$ and the discussion %l.406
is equal to the case with power of two modulus presented in Sec.\ %l.407
\ref{param}. We find %l.408
\begin{equation}%l.409 ú$
\NX =\left( P-1\right) P^{d-1}\hspace{.6ex},\hspace{2ex}Q_{1}=%l.410 ú$
\sqrt{2} \left( P-1\right) /P\hspace{.6ex},\hspace{2ex}Q_{n\ge %l.411 ó,ò...
2}\approx P^{1/n-1}\hspace{.6ex}.%l.412 ú$
\end{equation} %l.413
From a mathematical point of view odd prime number moduli give good %l.414
random number generators. In particular $ \tilde{N}_{1}=P-1$, $ %l.415 $
\tilde{M}_{1}=P$ whereas for the mixed multiplicative generator %l.416
with a power of two modulus $M$ we will find $ \tilde{N}_{1}= %l.417 $
\tilde{M}_{1}=\sqrt{2} M/4$. So we need $ M>4P/\sqrt{2}$ to obtain %l.418
power of two generators which behave better than generators with %l.419
prime number moduli. However, one has to take into account that %l.420
computers calculate automatically modulo powers of two. Moreover, %l.421
the power of two generator will be improved in Sec.\ \ref{int2} %l.422
until we achieve $ Q_{1}=1$. As a byproduct a better behavior of %l.423
$Q$$_{n}$ for $ n\ge 2$ is obtained, too.%l.424

Best performance allow prime numbers of the form $ P=2^{k}\pm 1$ %l.426
\cite{Knu}. In this case $ a\cdot b=c_{1}2^{k}+c_{2}$ leads to $ a%l.427 $
\cdot b=c_{2}\mp c_{1}{\rm\hspace{.38ex}mod\hspace{.38ex}}P$. The %l.428
extra effort, compared with a calculation $ {\rm\hspace%l.429 `...
{.38ex}mod\hspace{.38ex}}2^{k}$ is one addition and, which is more %l.430
important, the calculation of $c$$_{1}$. In Ex.\ 7.1 we discuss %l.431
the improved generator with $ M=2^{256}$ which can most easily be %l.432
changed to $ M=2^{128}$. Alternatively one may construct a generator %l.433
$ {\rm\hspace{.38ex}mod\hspace{.38ex}}2^{127}-1$ which is a prime %l.434
number. This generator will however be more time consuming and moreover %l.435
it has worse quality $ \tilde{N}_{1}\approx 2^{127}$, $ \tilde%l.436 $
{N}_{2}\approx 2^{63.5}$, $ \tilde{N}_{3}\approx 2^{42.3}$, etc.\ %l.437
vs.\ $ \tilde{N}_{1}=2^{129}$, $ \tilde{N}_{2}\approx 2^{86.3}$, %l.438
$ \tilde{N}_{3}\approx 2^{65}$, etc.\ for the generator with power %l.439
of two modulus (cf.\ Sec.\ \ref{int2}, Sec.\ \ref{param}). So, power %l.440
of two generators are more efficient than multiplicative generators %l.441
with prime number modulus. The situation is slightly different if %l.442
one considers multiply recursive generators with prime number modulus, %l.443
analyzed in Sec.\ \ref{multrec} and Ex.\ 7.2.%l.444

Multiplicative generators with prime number modulus and primitive %l.446
$a$ produce every random number $ \neq 0{\rm\hspace{.38ex}mod\hspace{.38ex}}P%l.447 $
$ exactly once in a period. We recommend to use a prime number modulus %l.448
only if one needs this quality. %l.449
\subsection{$ X_{0}=0$, $ X_{k+1}=aX_{k}+c{\rm\hspace{.38ex}mod\hspace{.38ex}}M%l.450 $
$\label{mm}} %l.451
Let $M$ be any non-prime modulus, $ {\rm\hspace{.38ex}gcd}(c,M)=1%l.452 $
$ and let $ a\neq 1$ have the following properties %l.453
\begin{equation}%l.454 ú$
\label{10}\hbox{\hspace{.38ex}1. }b\equiv {\rm\hspace{.38ex}gcd}%l.455 (,)...
\left( a-1,M\right) \hbox{ contains every prime factor of $M$,\hspace*{2cm}2. %l.456 "...
}4|b\hbox{ if }4|M.%l.457 ú$
\end{equation} %l.458
It was shown by Greenberger \cite{Gre} for $ M=2^d$ and by Hull and %l.459
Dobell \cite{Hul} for general $M$ that this leads to the quality %l.460
that every random number occurs exactly once in a period. This theorem %l.461
can also be obtained by harmonic analysis in a little more general %l.462
framework \cite{Schnetz}.%l.463

The generator is called mixed multiplicative generator with full %l.465
period.%l.466

\pagebreak[3]%l.468

\noindent {\bf Proposition} 2.\  %l.470
Let $M$$_{1}$ be a divisor of $M$, then %l.471
\begin{equation}%l.472 ú$
X_{kM_{1}}=ckM_{1}\cdot \left\{ %l.473 úBA...
\begin{array}{cl}1&\hbox{\hspace{.38ex}, $M$$_{1}$ odd\hspace{.38ex}}%l.474 úBA...
\\ %l.475 úBA...
1+b/2&\hbox{\hspace{.38ex}, $M$$_{1}$ odd\hspace{.38ex}}%l.476 úBA...
\end{array}\right\} {\rm\hspace{.38ex}mod\hspace{.38ex}}bM_{1}%l.477 ú$
\hspace{.6ex}.%l.478 ú$
\end{equation}%l.479

\pagebreak[3]%l.481

\noindent {\bf Proof}. With $ a\equiv db+1$ we find %l.483
\begin{displaymath}%l.484 ú$*
X_{kM_{1}}=c\frac{a^{kM_{1}}-1}{a-1} =ckM_{1}\left( 1+\frac{kM_{%l.485 ó,ò...
1}-1}{2}db\right) +cdb\sum _{j=3}^{M_{1}k}{M_{1}k\choose j}%l.486 (,)...
\left( db\right) ^{j-2}\hspace{.6ex}.%l.487 ú$*
\end{displaymath} %l.488
Obviously $ {\ErgoBbb N}\ni {M_{1}k\choose j}={M_{1}k-1\choose j%l.489 (\choose)...
-1}\frac{M_{1}k}{j}$. Since $j$ has at most $ j-2$ prime factors %l.490
for $ j\ge 3$ and $b$ has by definition every prime factor of $M$%l.491
$_{1}$ we obtain $ {\ErgoBbb N}\ni {M_{1}k-1\choose j-1}(db)^{j%l.492 ó,ò...
-2}k/j={M_{1}k\choose j}(db)^{j-2}{} /M_{1}$. Therefore the latter %l.493
term drops $ {\rm\hspace{.38ex}mod\hspace{.38ex}}bM_{0}$ and the %l.494
former gives the result.\hfill $\Box $%l.495
\pagebreak[3] %l.496

\pagebreak[3]%l.498

\noindent {\bf Theorem} 3.\  %l.500
Let $ s_{a,M/b}\equiv {\rm\hspace{.38ex}gcd}(s_{a},M/b)$. The Fourier %l.501
transform of the mixed multiplicative generator with full period %l.502
is %l.503
\begin{equation}%l.504 ú$
\label{46}|\hat{g}|^{2}\left( s_{0},{\bf s}\right) =bs_{a,M/b}%l.505 ú$
\delta _{s_{0}+cs_{a}=\frac{1}{2}{bs_{a,M/b}}\delta _{2|\frac{M%l.506 ó,ò...
}{bs_{a,M/b}{}} }{\rm\hspace{.38ex}mod\hspace{.38ex}}bs_{a,M/b}%l.507 ú$
}\hspace{.6ex},%l.508 ú$
\end{equation} %l.509
where the Kronecker $\delta $ gives 1 if the equation in the argument %l.510
holds and 0 otherwise.%l.511

\pagebreak[3]%l.513

\noindent {\bf Proof}. Since $ (c(a^{k}-1)/(a-1))_{k\in %l.515 { $
{\ErgoBbb Z}_M}$ gives every number $ {\rm\hspace{.38ex}mod\hspace{.38ex}}M%l.516 $
$ exactly once, in $ (a^{k}-1)_{k\in {\ErgoBbb Z}_M}{\rm\hspace%l.517 `...
{.38ex}mod\hspace{.38ex}}M$ every multiple of $b$ occurs $b$ times. %l.518
With $ {\bf s}\cdot ({\bf X}_{\Delta k+k}-{\bf X}_{k})=s_{a}a^{%l.519 ó,ò...
k}X_{\Delta k}$ we get from Eq.\ (\ref{15}) after a rearrangement %l.520
of the $k$-sum %l.521
\begin{displaymath}%l.522 ú$*
|\hat{g}|^{2}\left( s_{0},{\bf s}\right) =\frac{1}{M} \sum _{%l.523 { ú$*
\Delta k\in {\ErgoBbb Z}_M}\exp\left( \frac{2\pi i}{M} s_{0}%l.524 (,)...
\Delta k\right) \cdot b\sum _{k\in {\ErgoBbb Z}_{M/b}}\exp%l.525 (,)...
\left( \frac{2\pi i}{M} \left( kb+1\right) s_{a}X_{\Delta k}%l.526 ú$*
\right) \hspace{.6ex}.%l.527 ú$*
\end{displaymath} %l.528
The $k$-sum gives $ M/b$ if $ bs_{a}X_{\Delta k}=0{\rm\hspace%l.529 `...
{.38ex}mod\hspace{.38ex}}M$ and vanishes otherwise. Since $ {\rm\hspace%l.530 `...
{.38ex}gcd}(bs_{a},M)=bs_{a,M/b}$ only that $ \Delta k$ contribute %l.531
to $ |\hat{g}|^{2}$ for which $ X_{\Delta k}=0{\rm\hspace%l.532 `...
{.38ex}mod\hspace{.38ex}}M/bs_{a,M/b}$. The generator has a full %l.533
period, thus there are $ bs_{a,M/b}$ such $ \Delta k$. From Prop.\ %l.534
2 with $ M_{1}=M/bs_{a,M/b}$ we find that these have the form $ %l.535 $
\Delta k=kM_{1}$. Moreover, since $ M|s_{a}bM_{1}$, %l.536
\begin{eqnarray*}%l.537 ú$$*
|\hat{g}|^{2}\left( s_{0},{\bf s}\right) &=&\sum _{k\in %l.538 { ú$$*
{\ErgoBbb Z}_{bs_{a,M/b}}}\exp\left( \frac{2\pi i}{M} \left( s_{%l.539 ó,ò...
0}kM_{1}+s_{a}ckM_{1}\left( 1+\frac{b}{2}\delta _{2|M_{1}}%l.540 (,)...
\right) \right) \right) \\ %l.541 ú$$*
&=&bs_{a,M/b}\delta _{s_{0}+cs_{a}\left( 1+\frac{b}{2}\delta _{2%l.542 { (,)...
|M_{1}}\right) =0{\rm\hspace{.38ex}mod\hspace{.38ex}}bs_{a,M/b}%l.543 ú$$*
}\hspace{.6ex}.%l.544 ú$$*
\end{eqnarray*} %l.545
We get the result since $ cs_{a}b/2=-bs_{a,M/b}/2{\rm\hspace%l.546 `...
{.38ex}mod\hspace{.38ex}}bs_{a,M/b}$ if $ 2|M_{1}$.\hfill $\Box $%l.547
\pagebreak[3] %l.548

\vspace{1ex}%l.550
\noindent{}Now the proper choice of the parameters $a$ and $c$ can %l.551
be discussed. %l.552
\begin{enumerate}\item{}Choice of $c$. We can restrict ourselves %l.553 úBE
to $ 1\le c\le b/2$ since every $c$ emerges from a $ c\in (0,b/2%l.554 $...
)$ via translations ($ X_{k}\mapsto X_{k+\Delta k}-X_{\Delta k}%l.555 $...
=a^{\Delta k}X_{k}$) or reflection ($ X_{k}\mapsto -X_{k}$). We %l.556 úBE
can determine $c$ by the condition that there should be no small %l.557 úBE
$ (s_{0},s_{a})$, $ {\rm\hspace{.38ex}gcd}(M,s_{a})=1$ with $ s_{%l.558 ó,ò...
0}+s_{a}c=b/2\cdot \delta _{2|M/b}{\rm\hspace{.38ex}mod\hspace{.38ex}}b%l.559 $...
$. If $ M/b$ is odd $ c\approx \sqrt{b}$ gives $ Q_{1}(s_{0}%l.560 $...
\approx \sqrt{b} ,-1)\approx \sqrt{b+1} /b\approx b^{-1/2}$. In %l.561 úBE
the case where $ M/b$ is even and $b$ is small the choice $ c=1%l.562 $...
$ is best with the result $ Q_{1}(s_{0}\approx s_{1}\approx b/4%l.563 $...
)\approx \sqrt{2} \cdot (b/4)/b=\sqrt{2} /4\approx 0.35$.%l.564 úBE

In the case of Fig.\ 3 with the 'wrong' choice $ c=3$ one has $ a%l.566 $...
=9{\rm\hspace{.38ex}mod\hspace{.38ex}}16$, $ b=8$ and the smallest %l.567 úBE
$ (s_{0},s_{1})$ with non-vanishing $ \hat{g}$ is (1,1). Since $ %l.568 $...
|\hat{g}|^{2}(1,1)=8$ we obtain $ Q_{1}(1,1)=\sqrt{2} /8%l.569 $...
\approx 0.18$ (notice the correlations perpendicular to the (1,1)-direction %l.570 úBE
in Fig.\ 3I). With the right choice $ c=1$ (Fig.\ 4) it takes an %l.571 úBE
$ (s_{0},s_{1})=(1,3)$ (or (3,1)) to get $ |\hat{g}|^{2}=8$. Therefore %l.572 úBE
$ Q_{1}(1,3)=\sqrt{10} /8\approx 0.40$ which means that the random %l.573 úBE
numbers are more uniformly distributed in Fig.\ 4I. The large value %l.574 úBE
of $ Q_{1}(1,3)$ is yet misleading since $ Q_{1}=Q_{1}(-M/8,M/8%l.575 $...
)=\sqrt{2} /8$. However $ |\hat{g}|^{2}$ assumes the small value %l.576 úBE
of $Q$$_{1}$ at much less sites as in the case of $ c=3$ which means %l.577 úBE
that the choice $ c=1$ is better than $ c=3$.%l.578 úBE

Notice the similar pair distributions in Fig.\ 3II and Fig.\ 4II. %l.580 úBE
In general the quality dependence on $c$ can not be obtained by %l.581 úBE
the standard spectral test (corresponding to $ s_{0}=0$) since the %l.582 úBE
$n$-tuple distributions are only shifted by a change of $c$. %l.583 úBE
\item{}Choice of $b$. In general $b$ should be as small as possible %l.584 úBE
in order to prevent $ |\hat{g}|^{2}$ from being concentrated on %l.585 úBE
too few points. If $ M/b$ is odd, $ c\approx \sqrt{b}$ then $Q$$%l.586 ó,ò...
_{1}$ behaves like $ b^{-1/2}$. If $ M/b$ is even, $ c=1$ then $ Q%l.587 $...
_{1}(s_{0},s_{1})=Q_{1}(s_{0}\approx s_{1}\approx b/4)\approx %l.588 $...
\sqrt{2} /4$ for small $ (s_{0},s_{1})$ (cf.\ 1.). However $ Q_{%l.589 ó,ò...
1}(-M/b,M/b)=\sqrt{2} /b$ which forbids large values of $b$.%l.590 úBE

For a power of two modulus the smallest value possible is $ b=4$ %l.592 úBE
which implies $ Q_{1}=\sqrt{2} /4$ (Fig.\ 2). In particular $ M%l.593 $...
=10^d$ (Fig.\ 5) should be avoided since in this case $ b\ge 20%l.594 $...
$.%l.595 úBE

These arguments require $ s_{0}\neq 0$. They are not obtained by %l.597 úBE
the standard spectral test. %l.598 úBE
\item{}Choice of $a$. Up to now we have evaluated $ |\hat{g}|^{2%l.599 ó,ò...
}(s_{0},s_{1})$ for $ n=1$ which is given by $b$ and $c$. In order %l.600 úBE
to determine $a$ more precisely we have to look at the distribution %l.601 úBE
of $n$-tupels for $ n\ge 2$. In this case $ Q_{n}={\rm\hspace%l.602 `...
{.38ex}min\hspace{.38ex}}_{\bf s}Q_{n}(s_{0}=0,s_{a}=0)={\rm\hspace%l.603 `...
{.38ex}min\hspace{.38ex}}_{{\bf s}:s_{a}=0}|{\bf s}|/M$. A further %l.604 úBE
discussion of the choice of $a$ is postponed to Sec.\ \ref{param}. %l.605 úBE
We will see that a reasonable $a$ gives $ Q_{n}\approx M^{1/n-1%l.606 ó,ò...
}$.%l.607 úBE

Since $ s_{0}=0$ this part of the choice of $a$ is identical with %l.609 úBE
the standard spectral test.%l.610 úBE
\end{enumerate} %l.611
Let us summarize the result for $ M=2^d$, %l.612
\begin{eqnarray}%l.613 ú$$
&&c=1\hspace{.6ex},\hspace{2ex}a=5{\rm\hspace{.38ex}mod\hspace{.38ex}}8%l.614 ú$$
\hspace{.6ex},\hspace{2ex}{\rm\hspace{.38ex}max\hspace{.38ex}}_a%l.615 ú$$
{\rm min\hspace{.38ex}}_{{\bf s}:s_{a}=0}|{\bf s}|,\hbox{ for }n%l.616 ú$$
=2,3,{\ldots}\hbox{ gives\hspace{.38ex}}\\ %l.617 ú$$
&&\NX =M\hspace{.6ex},\hspace{2ex}Q_{1}=\sqrt{2} /4%l.618 ú$$
\hspace{.6ex},\hspace{2ex}Q_{n\ge 2}\approx M^{1/n-1}%l.619 ú$$
\hspace{.6ex}.%l.620 ú$$
\end{eqnarray} %l.621
A loss of randomness for $n$-tupels is avoided if one takes $ {\rm\hspace%l.622 `...
{.38ex}gcd}(n,M)=1$.%l.623

\subsection{$ X_{0}=0$, $ X_{k+1}=aX_{k}+c{\rm\hspace{.38ex}int}%l.625 $
(k/2){\rm\hspace{.38ex}mod\hspace{.38ex}}M$\label{int2}} %l.626
Let $ M=2^d$, $ 1\neq a$, $ a=1{\rm\hspace{.38ex}mod\hspace{.38ex}}4%l.627 $
$ and $c$ be odd (cf.\ Fig.\ 7).%l.628

We have seen that the quality parameter $ Q_{1}$ is not greater than %l.630
$ \sqrt{2} /4$ for the mixed multiplicative generator with power %l.631
of two modulus. This results in correlations along certain lines %l.632
in the figures (cf.\ e.g.\ Fig.\ 2I). This does not mean that mixed %l.633
multiplicative generators can not be used if one takes large enough %l.634
moduli (cf.\ Fig.\ 6 and Sec.\ \ref{param}). Nevertheless it is %l.635
worth to look for a generator with behaves better. The generator %l.636
presented in this section can be algebraically analyzed and it has %l.637
$ Q_{1}=1$. The implementation presented in Ex.\ 7.1 shows that %l.638
it has good performance. It is possible to motivate the generator %l.639
by geometrical arguments \cite{Schnetz}.%l.640

\pagebreak[3]%l.642

\noindent {\bf Theorem} 4.\ Let $ s_{a,M}={\rm\hspace{.38ex}gcd}%l.644 $
(s_{a},M)$. %l.645
\begin{equation}%l.646 ú$
\label{43}|\hat{g}|^{2}\left( s_{0},{\bf s}\right) =\delta _{s_{%l.647 ó,ò...
0}+{\bf s}\cdot {\bf X}_{0}^{\rm mm}=0{\rm\hspace{.38ex}mod\hspace{.38ex}}s_%l.648 2{ ú$
{a,M}}\cdot \left\{ %l.649 úBA...
\begin{array}{l}s_{a,M}\hspace*{6cm}\hbox{\hspace{.38ex}if }s_{a%l.650 { úBA...
,M}\neq M\\ %l.651 úBA...
M\left( 1+\cos\left( \frac{\pi }{M} \left( s_{0}+2c\sum \limits _{%l.652 ó,ò...
j=3}^{n}\frac{a^{j-1}-a^{\delta _{2|j}}}{a^{2}-1} s_{j}\right) %l.653 (,)...
\right) \right) \hbox{ else.\hspace{.38ex}}%l.654 úBA...
\end{array}\right. %l.655 ú$
\end{equation} %l.656
$ X_{k}^{\rm mm}=c(a^{k}-1)/(a-1)$ is the mixed multiplicative generator %l.657
related to $X$. The period is $ \NX =2M$.%l.658

It is useful to put the proof in a more general context. It is given %l.660
in \cite{Schnetz}. Here we only discuss the result. We obtain %l.661
\begin{equation}%l.662 ú$
\NX =2M\hspace{.6ex},\hspace{2ex}Q_{1}=1%l.663 ú$
\hspace{.6ex},\hspace{2ex}Q_{2}\approx M^{-1/3}%l.664 ú$
\hspace{.6ex},\hspace{2ex}Q_{n\ge 3}\approx M^{1/\left( n-1%l.665 { ú$
\right) -1}%l.666 ú$
\end{equation} %l.667
as will be shown in Sec.\ \ref{param}. It is advantageous to have %l.668
two parameters $a$ and $c$ at hand to optimize the quality of higher %l.669
$n$-tuples and not only $a$ as in the case of the mixed multiplicative %l.670
generator.%l.671

The generator does not provide full periods since $ |\hat{g}|^{2%l.673 ó,ò...
}(0,s_{1})=s_{1,M}\neq 2M\delta _{s_{1}=0}$. However the deviation %l.674
from an exact uniform distribution is not larger than in a finite %l.675
true random sequence. If one does not use the entire period of the %l.676
generator (and this is not recommended because of the $n$-tupel %l.677
distribution) the feature of having a full period is anyway irrelevant. %l.678
If, for some reasons, one insists in a full period we recommend %l.679
to use a multiplicative generator with prime number modulus or the %l.680
multiply recursive generator which will be analyzed next.%l.681

The generator of this section behaves in every aspect better than %l.683
the widely used mixed multiplicative generator. This is also confirmed %l.684
by the figures (cf.\ Fig.\ 2 and Fig.\ 7). The extra effort in calculating %l.685
random numbers is little (cf.\ Ex.\ 7.1). If $n$-tuples are used %l.686
one should take odd $n$ and occasionally omit one random number.%l.687

Nevertheless the most essential step for producing good random numbers %l.689
is to use large moduli (cf.\ Fig.\ 6 and Sec.\ \ref{param}).%l.690

\subsection{$ X_{0}\!=\!1,X_{-1}\!=\!. ..\!=\!X_{-r+1}\!=\!0$, $ X%l.692 $
_{k+1}\!=\!a_{r-1}X_{k}+. ..+a_{0}X_{k-r+1}{\rm\hspace{.38ex}mod\hspace{.38ex}}P%l.693 $
$\label{multrec}} %l.694
Let $P$ be a prime number and $X$$_{k}$ have the maximum period of %l.695
$ P^r-1$ (Fig.\ 8). In this case the random number generator has %l.696
a full period in the sense that every $r$-tuple $ (X_{0},%l.697 $
{\ldots},X_{r-1})\neq (0,{\ldots},0)$ occurs exactly once in a period.%l.698

\pagebreak[3]%l.700

\noindent {\bf Theorem} 5 (Grube \cite{Grube}). Let %l.702
\begin{equation}%l.703 ú$
P\left( \lambda \right) =\lambda ^r-a_{r-1}\lambda ^{r-1}-%l.704 ú$
{\ldots}-a_{0}.%l.705 ú$
\end{equation} %l.706
The corresponding generator has maximum period if and only if $P$ %l.707
is a primitive polynomial over $ {\ErgoBbb Z}_{P^r}$.%l.708

The proof is found in \cite[Satz 2.1]{Grube}.%l.710

\pagebreak[3]%l.712

\noindent {\bf Theorem} 6.\ Let $X$$_{k}$ be defined as above and %l.714
$ {\bf s}_{a}\equiv ({\bf s}\cdot {\bf X}_{k})_{0\le k<r}=(%l.715 $
\sum _{j=1}^{n}s_{j}X_{k+j-1}\hspace{-1pt})_{0\le k<r}$ then $ %l.716 $
|\hat{g}|^{2}$ is given by the following table. %l.717
\begin{equation}%l.718 ú$
\label{59}%l.719 úBA...
\begin{array}{c|cc}|\hat{g}|^{2}\left( s_{0},{\bf s}\right) &s_{%l.720 ó,ò...
0}=0&s_{0}\neq 0\\\hline  %l.721 úBA...
{\bf s}_{a}={\bf 0}&P^r-1&0\\ %l.722 úBA...
{\bf s}_{a}\neq {\bf 0}&1/\left( P^r-1\right) &P^r/\left( P^r-1%l.723 úBA...
\right) %l.724 úBA...
\end{array}%l.725 ú$
\end{equation}%l.726

\pagebreak[3]%l.728

\noindent {\bf Proof}. First we notice that $ ({\bf s}\cdot %l.730 $
{\bf X}_{k})_{k}$ obeys the same recursion relation as $ (X_{k}%l.731 $
)_{k}$ since $ {\bf s}\cdot {\bf X}_{k+1}=\sum _{j=1}^{n}s_{j}X_{%l.732 ó,ò...
k+j}=\sum _{\ell =1}^ra_{r-\ell }\sum _{j=1}^{n}s_{j}X_{k+j-%l.733 ó,ò...
\ell }=\sum _{\ell =1}^ra_{r-\ell }\,{\bf s}\cdot {\bf X}_{k+1-%l.734 ó,ò...
\ell }$. So, $ ({\bf s}\cdot {\bf X}_{k})_{k}$ is either identically %l.735
zero or it has maximum period. In the latter case every number $ %l.736 $
\in {\ErgoBbb Z}_P^\times $ is produced $ P^{r-1}$ times in a period %l.737
and the zero is generated $ P^{r-1}-1$ times. Since the same holds %l.738
for $ ({\bf s}\cdot ({\bf X}_{k+\Delta k}-{\bf X}_{k}))_{k}$ we %l.739
get %l.740
\begin{displaymath}%l.741 ú$*
\sum _{k\in {\ErgoBbb Z}_{P^r-1}}\exp\left( \frac{2\pi i}{P} %l.742 (,)...
{\bf s}\cdot \left( {\bf X}_{k+\Delta k}-{\bf X}_{k}\right) %l.743 ú$*
\right) =P^r\delta _{{\bf s}\cdot \left( {\bf X}_{k+\Delta k}-%l.744 (,)...
{\bf X}_{k}\right) =0\hspace*{1ex}\forall 0\le k<r}-1%l.745 ú$*
\hspace{.6ex}.%l.746 ú$*
\end{displaymath} %l.747
If $ {\bf s}_{a}={\bf 0}$ then $ {\bf s}\cdot {\bf X}_{k}=0$ $ %l.748 $
\forall k$, the Kronecker $\delta $ gives 1 and from Eq.\ (\ref{15}) %l.749
we obtain $ |\hat{g}|^{2}(s_{0},{\bf s}_{a}={\bf 0})=(P^r-1)%l.750 $
\delta _{s_{0}=0}$. If on the other hand $ {\bf s}_{a}\neq %l.751 $
{\bf 0}$ then $ ({\bf s}\cdot {\bf X}_{k})_{k}$ has maximum period %l.752
and the Kronecker $\delta $ vanishes unless $ \Delta k=0$. In this %l.753
case Eq.\ (\ref{15}) yields $ P^r/(P^r-1)-\delta _{s_{0}=0}$.\hfill $\Box $%l.754
\pagebreak[3] %l.755

The choice of parameters is determined by avoiding small $s$ with %l.757
$ {\bf s}_{a}={\bf 0}$. For practical purposes it is more convenient %l.758
to replace the condition $ {\bf s}_{a}=({\bf s}\cdot {\bf X}_{k%l.759 ó,ò...
})_{0\le k<r}={\bf 0}$ by the equivalent requirement $ {\bf 0}=%l.760 $
({\bf s}\cdot {\bf X}_{1-k})_{1\le k\le r}\Leftrightarrow \sum _{%l.761 ó,ò...
j=k}^{n}s_{j}X_{j-k}=0{\rm\hspace{.38ex}mod\hspace{.38ex}}P\hbox%l.762 "...
{\hspace{.38ex}, }k=1,2,{\ldots},{\rm\hspace{.38ex}min}(r,n)$. If %l.763
$ n\le r$ the only solution is $ {\bf s}={\bf 0}{\rm\hspace%l.764 `...
{.38ex}mod\hspace{.38ex}}P$. For $ n>r$ one has the problem of finding %l.765
the smallest lattice vector of an $n$-dimensional lattice. The unit %l.766
cell of this lattice has the volume $ P^r$ (cf.\ Sec.\ \ref{param}). %l.767
Thus for proper parameters the quality of the generator is %l.768
\begin{equation}%l.769 ú$
\NX =P^r-1\hspace{.6ex},\hspace{2ex}Q_{1}=Q_{2}={\ldots}=Q_r=%l.770 ú$
\sqrt{2} /\left( 1-P^{-r}\right) \hspace{.6ex},\hspace{2ex}Q_{n%l.771 { ú$
>r}\approx P^{r/n-r}\hspace{.6ex}.%l.772 ú$
\end{equation} %l.773
This is the first generator which has $ Q_{n}\ge 1$ for $ 1\le n%l.774 $
\le r>1$. For $ 2\le n\le r$ the generator has higher $ \tilde%l.775 $
{N}_{n}=P^r-1$ but lower $ \tilde{M}_{n}=P$ than the multiplicative %l.776
generator $ {\rm\hspace{.38ex}mod\hspace{.38ex}}P^r$ (Sec.\ \ref{mul}, %l.777
with $ \tilde{N}_{n}\approx  \tilde{M}_{n}\approx P^{r/n}$).%l.778

In particular if one needs the full periods this generator may be %l.780
recommended. For prime numbers of the form $ P=2^{k}\pm 1$ the generator %l.781
has good performance, too. If the prime factors of $ P^r-1$ are %l.782
known it is no problem to find multipliers which lead to a full %l.783
period. A short discussion of the choice of parameters for large %l.784
$P$ and $r$ is given in the next section and an implementation is %l.785
presented in Ex.\ 7.2.%l.786

\section{Choice of parameters\label{param}} %l.788
We start with a discussion of the mixed multiplicative generator %l.789
(the multiplicative generator is analogous). For practical purposes %l.790
we can restrict ourselves to $ M=2^d$ and $ a=5{\rm\hspace%l.791 `...
{.38ex}mod\hspace{.38ex}}8$. We set $ c=1$ which is equivalent to %l.792
any other odd $c$ and assume $ n\ge 2$ since the case $ n=1$ depends %l.793
only on $b$ which is 4 for $ a=5{\rm\hspace{.38ex}mod\hspace{.38ex}}8%l.794 $
$.%l.795

From Eq.\ (\ref{46}) we obtain, as long as $ bs_{a,M/b}<M$, that %l.797
$ \hat{g}$ vanishes unless $ s_{a,M/b}|s_{0}\neq 0$ and therefore %l.798
(A) $ Q_{n}(s_{0},{\bf s})=\sqrt{1+{\bf s}^{2}{} /s_{a,M/b}^{2}%l.799 ó,ò...
} /4>1/4$. However if $ s_{a,M/b}=M/b$ we get $ |\hat{g}|^{2}=M%l.800 $
$ for (B) $ s_{0}=s_{a}=0{\rm\hspace{.38ex}mod\hspace{.38ex}}M$. %l.801
This leads to $ Q_{n}=|{\bf s}|/M$ which for some $ {\bf s}$ is %l.802
much smaller than $ 1/4$.%l.803

So Eq.\ (B) is more important. We solve it for $s$$_{1}$ yielding %l.805
$ s_{1}=kM-as_{2}-a^{2}s_{3}-{\ldots}-a^{n-1}s_{n}$ depending on %l.806
the free integer constants $k$, $s$$_{2}$, $s$$_{3}$, {\dots}, $s$%l.807
$_{n}$ which give rise to an $n$-dimensional lattice (cf.\ \cite{Knu}). %l.808
The lattice is given by an $n$ by $n$ matrix $A$ according to $ %l.809 $
{\bf s}=A\cdot (k,s_{2},{\ldots},s_{n})^T$, and we read off %l.810
\begin{equation}%l.811 ú$
\label{48}A=\left( %l.812 úBA...
\begin{array} {ccccc}M&-a&-a^{2}&{\ldots}&-a^{n-1}{}\\ %l.813 úBA...
&1&&&\\ %l.814 úBA...
&&1&&\\ %l.815 úBA...
&&&\raisebox{ 3mm}{$\ddots$}&1%l.816 úBA...
\end{array}\right) \sim \left( %l.817 úBA...
\begin{array} {ccccc}M&-a&&&\\ %l.818 úBA...
&1&-a&&\\ %l.819 úBA...
&&1&\ddots &-a\\ %l.820 úBA...
&&&&1%l.821 úBA...
\end{array}\right) \hspace{.6ex},%l.822 ú$
\end{equation} %l.823
where zeros have been omitted and both matrices define the same lattice %l.824
since they differ only by $ SL(n,{\ErgoBbb Z})$ lattice transformations.%l.825

We denote the length of the smallest non-vanishing lattice vector %l.827
by $\nu _{n}$. Since the quality of the random numbers is determined %l.828
by $ \nu _{n}=MQ_{n}$ we search for an $a$ which large $\nu _{n%l.829 ó,ò
}$. Most important are small $n$, in particular the pair correlation %l.830
$ n=2$. In the best case the lattice has a cubic unit-cell and $%l.831
\nu _{n}$ is determined by the dimension of the lattice and the %l.832
volume of the unit-cell. Since the volume is given by the determinant %l.833
of $A$ we get as an approximate upper bound $ \nu _{n}%l.834 $
\lessapprox M^{1/n}$. The calculation of $\nu _{n}$ is a standard %l.835
problem in mathematics for which efficient algorithms exist \cite{Lenstra}.%l.836

To simplify the search for reasonable multipliers it is useful to %l.838
have also a lower bound for $\nu _{n}$. Due to the specific form %l.839
of $A$ it is easy to see that $\nu _{n}$ has to be larger than the %l.840
smallest ratio $ >1$ between two elements of then set $ \{1,a,a^{%l.841 ó,ò...
2}{\rm\hspace{.38ex}mod\hspace{.38ex}}M,{\ldots},a^{n-1}{\rm\hspace%l.842 `...
{.38ex}mod\hspace{.38ex}}M,M\}$. If we take e.g.\ $ a\approx M^{%l.843 ó,ò...
1/2}$ we find $ \nu _{2}\gtrapprox M^{1/2}$ which is identical with %l.844
the upper bound.%l.845

Similarly we obtain $ \nu _{3}\approx M^{1/3}$ if we take $ a%l.847 $
\approx M^{1/3}$ or $ a\approx M^{2/3}$. However this is not compatible %l.848
with $ a\approx M^{1/2}$ and we only get $ \nu _{2}\gtrapprox M^{%l.849 ó,ò...
1/3}$. On the other hand we can take $ a\approx M^{1/2}{} +%l.850 ó,ò...
\frac{1}{2}M^{1/4}$ which differs little from $M$$^{1/2}$. Therefore %l.851
$ \nu _{2}\approx M^{1/2}$ and since $ a^{2}\approx M+M^{3/4}{} %l.852 $
+\frac{1}{4}M^{1/4}\approx M^{3/4}{\rm\hspace{.38ex}mod\hspace{.38ex}}M%l.853 $
$ we have $ \nu _{3}\gtrapprox M^{1/4}$. Generally, with $ a%l.854 $
\approx M^{1/2}{} +\frac{1}{2}M^{1/4}{} +{\ldots}+\frac{1}{k-1}M^%l.855 { $
{1/2^{k-1}}$ (the plus signs may as well be replaced by minus signs) %l.856
we get $ \nu _{n}\gtrapprox M^{1/2^{n-1}}$ as long as $ k\ge n$ %l.857
and $ M^{1/2^{n-1}}\gg 1$. Note that $ M^{1/2^{n-1}}$ is only a %l.858
lower bound for $\nu _{n}$. In the generic case $\nu _{n}$ will %l.859
be close to $M$$^{1/n}$ (cf.\ Ex.\ 7.1).%l.860

Obviously $\nu _{n}$ increases with $M$. For all practical purposes %l.862
the magnitude of $M$ is only limited by the performance of the generator. %l.863
In practice one has to split $M$ into groups of digits (16 or 32 %l.864
bit) that can be treated on a computer. The multiplication by $a$ %l.865
performs best if the pre-factors $ 1/j$ are omitted. This should %l.866
be done even though for $ a\approx M^{1/2}+M^{1/4}+{\ldots}+M^{1%l.867 { $
/2^{k-1}}$ the lower bounds for $\nu _{n}$ decrease, $ \nu _{n}%l.868 $
\gtrapprox M^{1/2^{n-1}}/(n-1)!$. Note that the number of digits %l.869
of $M$ is much more important for randomness than the fine-tuning %l.870
of $a$.%l.871

Finally, we have to add not too small a constant $ a_{0}=5{\rm\hspace%l.873 `...
{.38ex}mod\hspace{.38ex}}8$ (16 or 32 bit) to the sum of powers %l.874
of $M$. This constant can be fixed by explicit calculation of the %l.875
$\nu _{n}$ or by looking at (A) from the beginning of this section %l.876
which implies that $a$$_{0}$ should have large $ |{\bf s}|$ for %l.877
all $ 16<m=s_{a,M}|M^{1/2^{k-1}}$. A suitable choice is e.g.\ $ a%l.878 $
_{0}=3~580~621~541=62~181{\rm\hspace{.38ex}mod\hspace{.38ex}}2^{%l.879 ó,ò...
16}$. With this value of $a$$_{0}$ we find $ |{\bf s}|\approx m^{%l.880 ó,ò...
1/n}$ for $ n=2,3$.%l.881

We summarize the result for the parameters of the mixed multiplicative %l.883
generator: %l.884
\begin{eqnarray}%l.885 ú$$
&&M=2^{2^{k}d_{0}}\hspace{.6ex},\hspace{2ex}c=1%l.886 ú$$
\hspace{.6ex},\hspace{2ex}a=2^{2^{k-1}d_{0}}+2^{2^{k-2}d_{0}}+%l.887 ú$$
{\ldots}+2^{2d_{0}}+a_{0}\hspace{.6ex},\hspace{2ex}\hbox%l.888 "...
{\hspace{.38ex}with\hspace{.38ex}}\nonumber \\ %l.889 ú$$
&&a_{0}=5{\rm\hspace{.38ex}mod\hspace{.38ex}}8%l.890 ú$$
\hspace{.6ex},\hspace{2ex}a_{0}\approx 2^{d_{0}}%l.891 ú$$
\hspace{.6ex},\hspace{2ex}\hbox{\hspace{.38ex}e.g. }a_{0}=3~580~621%l.892 ú$$
~541{\rm\hspace{.38ex}mod\hspace{.38ex}}2^{d_{0}}\hbox%l.893 "...
{\hspace{.38ex}, leads to\hspace{.38ex}}\nonumber \\ %l.894 ú$$
&&\label{49}\NX =2^{256}\hspace{.6ex},\hspace{2ex}Q_{1}=\sqrt{2%l.895 ...
} /4\hspace{.6ex},\hspace{2ex}Q_{n}\approx M^{1/n-1}%l.896 ú$$
\hspace{.6ex}.%l.897 ú$$
\end{eqnarray}%l.898

Now we turn to the improved generator of Sec.\ \ref{int2}. The Fourier %l.900
transform of the generator is given by Eq.\ (\ref{43}). We set $ n%l.901 $
\ge 2$ since independently of the parameters $ Q_{1}=1$. Further %l.902
on, we fix an $ m|M$ and find that $ |\hat{g}|^{2}=m$ if and only %l.903
if (C) $ s_{a}=km$, $k$ odd if $ m<M$, and (D) $ s_{0}+\frac{c}%l.904 ...
{a-1} \sum _{j=2}^{n}(a^{j-1}-1)s_{j}=\ell m$. (We neglect here %l.905
that $|$\^{g}$|^{2}$ may even be $ 2M$ for $ m=M$.) Eq.\ (C) can %l.906
be solved for $s$$_{1}$ and Eq.\ (D) for $s$$_{0}$ depending on %l.907
the integer parameters $k$, $ \ell $, $s$$_{2}$, {\dots}, $s$$%l.908 ó,ò
_{n}$. This gives rise to an ($ n+1$)-dimensional lattice (for $ m%l.909 $
<M$ we actually obtain an affine sub-lattice since $k$ has to be %l.910
odd) determined by the matrix $B$ via $ (s_{0},{\bf s})=B\cdot %l.911 $
(\ell ,k,s_{2},{\ldots},s_{n})^T$, %l.912
\begin{eqnarray}%l.913 ú$$
\hspace{-1ex}B&\hspace{-1ex}=&\hspace{-1ex}\left( %l.914 úBA...
\begin{array} {ccccc}m&&-c&{\ldots}&-c\left( a^{n-2}+{\ldots}+1%l.915 úBA...
\right) \\ %l.916 úBA...
&m&-a&{\ldots}&-a^{n-1}{}\\ %l.917 úBA...
&&1&&\\ %l.918 úBA...
&&&\raisebox{ 1mm}{$\ddots$}&\\ %l.919 úBA...
&&&&1%l.920 úBA...
\end{array}\right) \sim \left( %l.921 úBA...
\begin{array} {cccccc}m&&-c&-c&{\ldots}&-c\\ %l.922 úBA...
&m&-a&&&\\ %l.923 úBA...
&&1&-a&&\\ %l.924 úBA...
&&&1&\ddots &-a\\ %l.925 úBA...
&&&&&1%l.926 úBA...
\end{array}\right) \\ %l.927 ú$$
\label{47}&\hspace{-1ex}\sim &\hspace{-1ex}\left( %l.928 úBA...
\begin{array} {ccccccc}m&&-c&&&&\\ %l.929 úBA...
&m&-a&a&a^{2}+a&{\ldots}&a^{n-2}+{\ldots}+a\\ %l.930 úBA...
&&1&-a-1&-a^{2}-a-1&{\ldots}&-a^{n-2}-{\ldots}-1\\ %l.931 úBA...
&&&1&&&\\ %l.932 úBA...
&&&&1&&\\ %l.933 úBA...
&&&&&\raisebox{ 3mm}{$\ddots$}&1%l.934 úBA...
\end{array}\right) \hspace{.6ex},%l.935 ú$$
\end{eqnarray} %l.936
where again zeros have been omitted.%l.937

\noindent{}B describes an ($ n+1$)-dimensional lattice which has %l.939
a unit-cell with volume $m$$^{2}$. However this does not imply that %l.940
the smallest lattice vector $\nu _{n}$ has length of about $ m^%l.941 { $
{2/(n+1)}$. We see from (\ref{47}) that there exists an ($ n-1$)-dimensional %l.942
sub-lattice with $ s_{0}=0$ and $ s_{2}=-s_{1}-s_{3}-{\ldots}-s_{%l.943 ó,ò...
n}$ (delete the first and the third row and column in (\ref{47})). %l.944
The unit-cell of the sub-lattice has volume $m$ and $ \nu _{n}%l.945 $
\approx m^{1/(n-1)}$ which, for $ n\ge 4$, is smaller than $ m^%l.946 { $
{2/(n+1)}$. The smallest lattice vector for $ n\ge 4$ will have %l.947
the form $ (0,s_{1},-s_{1}-s_{3}-{\ldots}-s_{n},s_{3},{\ldots},s%l.948 $
_{n})$ with the length $ (s_{1}^{2}+s_{3}^{2}+{\ldots}+s_{n}^{2%l.949 ó,ò...
}+(s_{1}+s_{3}+{\ldots}+s_{n})^{2})^{1/2}$. Since this is of about %l.950
the same magnitude as $ (s_{1}^{2}+s_{3}^{2}+{\ldots}+s_{n}^{2}%l.951 $
)^{1/2}$ we may simply omit $s$$_{2}$ and reduce the problem to %l.952
the ($ n-1$) dimensions given by ($ s_{1},s_{3},{\ldots},s_{n}$). %l.953
Geometrically this means that the lattice corresponding to $B$ for %l.954
$ n\ge 4$ never has an approximately cubic unit-cell. Note moreover %l.955
that the sub-lattice is independent of $c$ which means that $c$ %l.956
can not be fixed by looking at the $n$-tupel distributions for $ n%l.957 $
\ge 4$.%l.958

The smallest value of $ Q_{n}=\nu _{n}/m$ is obtained for $ m=M$ %l.960
which is thus the most important case. For $ m=M$ we are not restricted %l.961
to odd $k$. The situation is similar to the ($ n-1$)-dimensional %l.962
case of the mixed multiplicative generator, Eq.\ (\ref{48}), with %l.963
$ -a^{j}$ replaced by $ a^{j}+a^{j-1}+{\ldots}+a$. This allows us %l.964
to use $ a=M^{1/2}+M^{1/4}+{\ldots}+M^{1/2^{k-1}}+a_{0}$ again. %l.965
Since $ 1\ll a\approx M^{1/2}\ll a^{2}{\rm\hspace{.38ex}mod\hspace{.38ex}}M%l.966 $
\approx 2M^{3/4}\ll {\ldots}\ll a^{n-2}{\rm\hspace{.38ex}mod\hspace{.38ex}}M%l.967 $
\approx (n-2)!M^{1-2^{2-n}}\ll M$ we have $ a^{j}+a^{j-1}+%l.968 $
{\ldots}+a\approx a^{j}{\rm\hspace{.38ex}mod\hspace{.38ex}}M$. The %l.969
minus sign is irrelevant, thus we can copy the corresponding lower %l.970
bounds from the mixed multiplicative generator: $ \nu _{n}%l.971 $
\gtrapprox M^{1/2^{n-2}}/(n-2)!$ for $ k+1\ge n\ge 4$. The constant %l.972
$a$$_{0}$ is given by the case $ m<M$ as will be discussed below.%l.973

The constant $c$ can be fixed by the case $ n=2$. We have to meet %l.975
two equations (E) $ s_{1}+as_{2}=0{\rm\hspace{.38ex}mod\hspace{.38ex}}M%l.976 $
$ and (F) $ s_{0}+cs_{2}=0{\rm\hspace{.38ex}mod\hspace{.38ex}}M%l.977 $
$ to get $ |\hat{g}|^{2}=M$. Both equations are solved by e.g.\ %l.978
$ s_{0}=-c$, $ s_{1}=-a\approx -M^{1/2}$, $ s_{2}=1$ with $ |(s_{%l.979 ó,ò...
0},{\bf s})|\approx (c^{2}+M)^{1/2}$. In order to reach the theoretical %l.980
limit $ \nu _{2}\approx M^{2/3}$ one needs $ c\gtrapprox M^{2/3%l.981 ó,ò...
}$. So, the simplest ansatz for $c$ is $ c=M^\lambda +1$ for $ %l.982 $
\lambda \ge 2/3$, $ M^\lambda \in {\ErgoBbb N}$. On the other hand, %l.983
if $ s_{1}=M^{1-\lambda }s_{1}'$, $ s_{2}=M^{1-\lambda }s_{2}'$ %l.984
then $ s_{1}'+as_{2}'=0{\rm\hspace{.38ex}mod\hspace{.38ex}}M^%l.985 $
\lambda $ has a solution with $ |{\bf s}'|\lessapprox M^{%l.986 { $
\lambda /2}$. Since (F) is solved by $ s_{0}=-M^{1-\lambda }s_{%l.987 ó,ò...
2}'$ we find $ |(s_{0},{\bf s})|\approx |s_{0}|\lessapprox M^{1%l.988 { $
-\lambda }M^{\lambda /2}=M^{1-\lambda /2}$. To allow for the maximum %l.989
value $ M^{2/3}$ one needs $ \lambda \le 2/3$. In general, $c$ should %l.990
not have more successive zero digits than $ M^{2/3}$ has. The simplest %l.991
reasonable choice is therefore $ c=M^{2/3}+1$. We can generalize %l.992
this slightly to $ c=(2^{d_{1}}+1)c_{0}$, where $c$$_{0}$ is a 16 %l.993
or 32 bit number and $ 2^{d_{1}}\le M^{2/3}\le c_{0}2^{d_{1}}$. %l.994
This choice of $c$ leads to best performance among all reasonable %l.995
$c$. We will see in Ex.\ 7.1 that it actually gives $ \nu _{2}%l.996 $
\approx M^{2/3}$ and $ \nu _{3}\approx M^{1/2}$. As a lower bound %l.997
for $\nu _{2}$, $\nu _{3}$ one has only the values $M$$^{1/2}$, %l.998
$M$$^{1/4}$ that are obtained from Eqs.\ (E), (C) alone.%l.999

Now we determine $c$$_{0}$ and $a$$_{0}$ by looking at $ s_{a,M}%l.1001 $
=m<M$. The case $ m<M$ is more important than for the mixed multiplicative %l.1002
generator since $Q$$_{n}$ is not limited by $ 1/4$. To some extent %l.1003
the smaller $Q$$_{n}$ for $ m<M$ is compensated by the fact that %l.1004
for small $m$ there are more points with $ s_{a}={\rm\hspace%l.1005 `...
{.38ex}odd\hspace{.38ex}}\cdot m$. We use $ a_{0}=3~580~621~541%l.1006 $
$ as for the mixed multiplicative generator and find with $ c_{%l.1007 ó,ò...
0}=3~370~134~727=11~463{\rm\hspace{.38ex}mod\hspace{.38ex}}2^{1%l.1008 ó,ò...
6}$ that $ Q_{2}(s_{0},{\bf s})\approx m^{2/3-1}$ and $ Q_{3}(s_{%l.1009 ó,ò...
0},{\bf s})\approx m^{1/2-1}$ if $ s_{a,M}=m$.%l.1010

We summarize the result for the generator of Sec.\ \ref{int2}: %l.1012
\begin{eqnarray}%l.1013 ú$$
&&M=2^{2^{k}d_{0}},\hspace*{1ex}a=2^{2^{k-1}d_{0}}+2^{2^{k-2}d_{%l.1014 ó,ò...
0}}+{\ldots}+2^{2d_{0}}+a_{0},\hspace*{1ex}c=\left( 2^{{\rm int}%l.1015 { (,)...
\left( 2^{k+1}{} /3\right) d_{0}}+1\right) c_{0},\hbox{ with\hspace{.38ex}}%l.1016 ú$$
\nonumber \\ %l.1017 ú$$
&&a_{0}=5{\rm\hspace{.38ex}mod\hspace{.38ex}}8%l.1018 ú$$
\hspace{.6ex},\hspace{2ex}a_{0}\approx 2^{d_{0}}%l.1019 ú$$
\hspace{.6ex},\hspace{2ex}c_{0}{\rm\hspace{.38ex}odd\hspace{.38ex}}%l.1020 ú$$
\hspace{.6ex},\hspace{2ex}c_{0}\ge 2^{2/3\cdot d_{0}}%l.1021 ú$$
\hspace{.6ex},\nonumber \\ %l.1022 ú$$
&&\hbox{\hspace{.38ex}e.g. }a_{0}=3~580~621~541{\rm\hspace%l.1023 `...
{.38ex}mod\hspace{.38ex}}2^{d_{0}}\hspace{.6ex},\hspace{2ex}c_{%l.1024 ó,ò...
0}=3~370~134~727{\rm\hspace{.38ex}mod\hspace{.38ex}}2^{d_{0}}\hbox%l.1025 "...
{\hspace{.38ex}\hspace*{2ex}leads to\hspace{.38ex}}\nonumber \\%l.1026 ú$$
 %l.1027 ú$$
&&\label{50}\NX =2^{257}\hspace{.6ex},\hspace{2ex}Q_{1}=1%l.1028 ú$$
\hspace{.6ex},\hspace{2ex}Q_{2}\approx M^{2/3-1}%l.1029 ú$$
\hspace{.6ex},\hspace{2ex}Q_{n\ge 3}\approx M^{1/\left( n-1%l.1030 { ú$$
\right) -1}\hspace{.6ex}.%l.1031 ú$$
\end{eqnarray}%l.1032

Finally we give a short discussion of the multiply recursive generator %l.1034
of Sec.\ \ref{multrec} (cf.\ Ex.\ 7.2).%l.1035

\noindent{}P should not be taken too small to provide enough digits %l.1037
for the random numbers. To optimize the performance one should use %l.1038
a prime number of the form $ P=2^d\pm 1$, e.g.\ $ P=2^{31}-1$. Moreover %l.1039
we set $ a_{r-1}=1$, $ a_{r-2}={\ldots}=a_{1}=0$.%l.1040

The most severe problem is to find the prime factors of $ P^r-1$. %l.1042
To this end it is useful to take $ r=2^k$ since in this case $ P^%l.1043 { $
{2^{k}}-1=(P^{2^{k-1}}+1)\cdot {\ldots}\cdot (P+1)\cdot (P-1)$ and %l.1044
one is basically left with the problem to determine the prime factors %l.1045
of $ P^{2^{k-1}}+1$.%l.1046

Afterwards it is easy to find an $ a_{0}\in {\ErgoBbb Z}_P^%l.1048 $
\times $ that makes the polynomial $ P(\lambda )=\lambda ^r-%l.1049 $
\lambda ^{r-1}-a_{0}$ primitive over $ {\ErgoBbb Z}_{P^r}$. Since %l.1050
\begin{equation}%l.1051 ú$
\label{60}\left( %l.1052 úBA...
\begin{array} {c}X_{k}{}\\ %l.1053 úBA...
X_{k-1}{}\\ %l.1054 úBA...
\vdots \\ %l.1055 úBA...
X_{k-r+1}%l.1056 úBA...
\end{array}\right) =X^{k}\cdot \left( %l.1057 úBA...
\begin{array}{c}1\\ %l.1058 úBA...
0\\ %l.1059 úBA...
\vdots \\ %l.1060 úBA...
0%l.1061 úBA...
\end{array}\right) \hspace{.6ex},\hspace{2ex}\hbox%l.1062 "...
{\hspace{.38ex}with }X\equiv \left( %l.1063 úBA...
\begin{array} {ccccc}a_{r-1}&a_{r-2}&{\ldots}&a_{1}&a_{0}{}\\ %l.1064 úBA...
1&0&&&\\ %l.1065 úBA...
&&\ddots &&\\ %l.1066 úBA...
&&&1&0%l.1067 úBA...
\end{array}\right) \hspace{.6ex},%l.1068 ú$
\end{equation} %l.1069
a necessary and sufficient condition for a maximum period is $ X^%l.1070 { $
{(P^r-1)}=1\hspace{-.6ex}{\rm l} {\rm\hspace{.38ex}mod\hspace{.38ex}}P%l.1071 $
$ and $ X^{(P^r-1)/p}\neq 1\hspace{-.6ex}{\rm l} {\rm\hspace%l.1072 `...
{.38ex}mod\hspace{.38ex}}P$ for all prime factors $p$ of $ P^r-1%l.1073 $
$. High powers of $X$ are easily computed. If $ N=\sum b_{i}2^{%l.1074 ó,ò...
i}$, $ b_{i}\in \{0,1\}$ then $ X^N=\prod _{\{i:b_{i}=1\}}X^{2^{%l.1075 ó,ò...
i}}$ and $ X^{2^{i}}=(X^{2^{i-1}})^{2}$.%l.1076

Now one has to check the $n$-tupel distributions for $ n>r$. We found %l.1078
(Eq.\ (\ref{59})) that $ |\hat{g}|^{2}=P^r-1$ if and only if $ s%l.1079 $
_{0}=0$ and $ {\bf s}_{a}={\bf 0}$. The latter equation is equivalent %l.1080
to $ 0=\sum _{j=k}^{n}s_{j}X_{j-k}=\ell _{k}P$, $ k=1,2,%l.1081 $
{\ldots},r$, $ \ell _{k}\in {\ErgoBbb Z}$ (cf.\ Sec.\ \ref{multrec}) %l.1082
and gives rise to an $n$-dimensional lattice determined by $C$ via %l.1083
$ {\bf s}=C\cdot (\ell _{1},{\ldots},\ell _r,s_{r+1},{\ldots},s_{%l.1084 ó,ò...
n})^T$, $ C=C_{1}\cdots C_r$, %l.1085
\begin{equation}%l.1086 ú$
C_{k}=\left( %l.1087 úBA...
\begin{array}{cccccc}1&&&&&\\ %l.1088 úBA...
&\raisebox{ 3mm}{$\ddots$}\hspace*{1ex}1&&&&\\ %l.1089 úBA...
&&P&-X_{1}&{\ldots}&-X_{n-k}{}\\ %l.1090 úBA...
&&&1&&\\ %l.1091 úBA...
&&&&\raisebox{ 3mm}{$\ddots$}&1%l.1092 úBA...
\end{array}\right) .\hspace*{1ex}C\sim C_{0}\equiv \left( %l.1093 úBA...
\begin{array}{cccccc} %l.1094 úBA...
P&&&-a_{0}&&\\ %l.1095 úBA...
&\raisebox{ 3mm}{$\ddots$}&P&-1&\raisebox{ 3mm}{$\ddots$}&-a_{0}%l.1096 úBA...
{}\\ %l.1097 úBA...
&&&1&\raisebox{ 3mm}{$\ddots$}&-1\\ %l.1098 úBA...
&&&&\raisebox{ 3mm}{$\ddots$}&1%l.1099 úBA...
\end{array}\right) %l.1100 ú$
\end{equation} %l.1101
(after some lattice transformations) if $ r<n\le 2r$ and $ a_{r-1%l.1102 $
}=1$, $ a_{r-2}={\ldots}=a_{1}=0$. The determinant of $C$$_{0}$ %l.1103
is $ P^r$, however the symmetry of $C$$_{0}$ leads to $ \nu _{r%l.1104 { $
+1}={\ldots}=\nu _{2r}\equiv \nu $ which is given by the shortest %l.1105
lattice vector of the 2 by 3 matrix $ %l.1106 $
\renewcommand{\arraystretch}{1}\left( %l.1107 úBA...
\begin{array} {ccc}P&0&0\\ %l.1108 úBA...
-a_{0}&-1&1%l.1109 úBA...
\end{array}\right) ^T\renewcommand{\arraystretch}{1.2}$. Since the %l.1110
second and third row are identical (up to a minus sign) the problem %l.1111
is analogous to the calculation of $\nu _{2}$ in the case of the %l.1112
mixed multiplicative generator. We obtain $ \nu \lessapprox 2^{%l.1113 ó,ò...
1/4}P^{1/2}$ with $ a_{0}\approx 2^{1/4}P^{1/2}\approx 55109$ for %l.1114
$ P=2^{31}-1$.%l.1115

We summarize the result for the generator of Sec.\ \ref{multrec}: %l.1117
\begin{eqnarray}%l.1118 ú$$
&&P=2^d-1,\hbox{ prime\hspace{.38ex}},\hspace*{1ex}r=2^{k},%l.1119 ú$$
\hspace*{1ex}a_{r-1}=1,\hspace*{1ex}a_{r-2}={\ldots}=a_{1}=0,%l.1120 ú$$
\hspace*{1ex}a_{0}\approx 2^{1/4}P^{1/2},\hbox{ with\hspace{.38ex}}%l.1121 ú$$
\nonumber \\ %l.1122 ú$$
&&X^{\left( P^r-1\right) }=1\hspace{-.6ex}{\rm l} {\rm\hspace%l.1123 `...
{.38ex}mod\hspace{.38ex}}P\hbox{ and }X^{\left( P^r-1\right) /p%l.1124 ú$$
}\neq 1\hspace{-.6ex}{\rm l} {\rm\hspace{.38ex}mod\hspace{.38ex}}P%l.1125 ú$$
\hspace*{1ex}\forall p|\left( P^r-1\right) ,\hspace*{1ex}p\hbox{ %l.1126 "...
prime, leads to\hspace{.38ex}}\nonumber \\ %l.1127 ú$$
&&\NX =\left( P^r-1\right) \hspace{.6ex},\hspace{2ex}Q_{1}=%l.1128 ú$$
{\ldots}=Q_r\approx \sqrt{2} \hspace{.6ex},\hspace{2ex}Q_{r+1}=%l.1129 ú$$
{\ldots}=Q_{2r}\approx 2^{1/4}P^{1/2-r}\hspace{.6ex}.%l.1130 ú$$
\end{eqnarray} %l.1131
Notice that the effort for calculating random numbers does not increase %l.1132
with $r$.%l.1133

Let us finally mention that the quality of the $n$-tupel $ %l.1135 $
{\bf X}_{\ell }$ of the (non-successive) random numbers $ X_{%l.1136 ó,ò...
\ell },X_{k_{2}+\ell },{\ldots},X_{k_{n}+\ell }$ deteriorates to %l.1137
$ Q_{n}\lessapprox P^{(r-d)/n-r}$ if there exist $ d>r-n$ values %l.1138
of $ j\in \{-1,{\ldots},-r\}$ with $ {\bf X}_{j}={\bf 0}$ (see the %l.1139
remark at the end of Sec.\ \ref{gentest}). In particular if $ X_{%l.1140 ó,ò...
k-1}={\ldots}=X_{k-r+1}=0$ the pair $ (X_{0},X_{k})$ has quality %l.1141
of less than $ P^{1/2-r}$ because $ aX_{\ell }=bX_{k+\ell }$ $ %l.1142 $
\forall \ell $ if $ a=bX_{k}{\rm\hspace{.38ex}mod\hspace{.38ex}}P%l.1143 $
$. From Eq.\ (\ref{60}) we see immediately that this happens for %l.1144
multiples of $ k=(P^r-1)/(P-1)$ (notice the equidistant zeros in %l.1145
Fig.\ 8I). This makes it not desirable to use more than $ (P^r-1%l.1146 $
)/(P-1)$ multiply recursive random numbers.%l.1147

\pagebreak[3]%l.1149

\noindent {\bf Example} 7.\  %l.1151
\begin{enumerate}\item{}We set $ M=2^{256}=2^{2^{4}\cdot 16}$, $ a%l.1152 $...
=2^{128}+2^{64}+2^{32}+62~181$ and in case of the generator of Sec.\ %l.1153 úBE
\ref{int2} $ c=(2^{160}+1)\cdot 11~463$. In the following table %l.1154 úBE
we compare the mixed multiplicative generator with the generator %l.1155 úBE
of Sec.\ \ref{int2}. The results can easily be obtained with a computer %l.1156 úBE
algebra program and Eq.\ (\ref{43}). %l.1157 úBE
\begin{equation}%l.1158 ú$...
\begin{array}{c|cc|cc} Q_{n}\equiv M^{\alpha _{n}-1}&X_{k+1}=aX_{%l.1159 ó,ò...
k}+1&\hbox{\hspace{.38ex}Eq.\ (\ref{49})\hspace{.38ex}}&X_{k+1}%l.1160 úBA...
=aX_{k}+c{\rm\hspace{.38ex}ink}\left( k/2\right) &\hbox%l.1161 "...
{\hspace{.38ex}Eq.\ (\ref{50})\hspace{.38ex}}\\ %l.1162 úBA...
\alpha _{1}&0.99414&0.99414&1.00000&1.00000\\ %l.1163 úBA...
\alpha _{2}&0.50000&0.50000&0.65658&0.66667\\ %l.1164 úBA...
\alpha _{3}&0.33203&0.33333&0.49783&0.50000\\ %l.1165 úBA...
\alpha _{4}&0.24859&0.25000&0.33436&0.33333\\ %l.1166 úBA...
\alpha _{5}&0.19721&0.20000&0.24636&0.25000\\ %l.1167 úBA...
\alpha _{6}&0.16335&0.16667&0.19882&0.20000%l.1168 úBA...
\end{array}%l.1169 ú$...
\end{equation} %l.1170 úBE
We see a good agreement of the quality parameters with the approximate %l.1171 úBE
upper bounds. This means that our choice of parameters is satisfactory. %l.1172 úBE
Moreover the table confirms that the quality parameter of the generator %l.1173 úBE
of Sec.\ \ref{int2} lies above the quality of the mixed multiplicative %l.1174 úBE
generator.%l.1175 úBE

Finally we present an implementation of the generator in Pascal. %l.1177 úBE
We group the digits of $X$$_{k}$ to 16 blocks of 16 digits {\tt %l.1178 { úBE
X[1], {\dots}, X[16]} starting from the highest digits.%l.1179 úBE

\vspace{1ex}%l.1181 úBE
\noindent{}\tt unit random1;\newline %l.1182 úBE
interface\newline %l.1183 úBE
const n=16; n0=(n+2) div 3; a0=62181; c0=11463;\newline %l.1184 úBE
var X:array[1..n] of longint;\newline %l.1185 úBE
procedure nextrandom;\newline %l.1186 úBE
implementation\newline %l.1187 úBE
var even:boolean; i:word; c:longint;\newline %l.1188 úBE
procedure nextrandom;\newline %l.1189 úBE
var j,k:word;\newline %l.1190 úBE
begin\newline %l.1191 úBE
if even then inc(c,c0); even:=not even;\newline %l.1192 úBE
for j:=1 to n do begin\newline %l.1193 úBE
\hspace*{1cm}X[j]:=X[j]*a0;\newline %l.1194 úBE
\hspace*{1cm}k:=2;while j+k$<$=n do begin inc(X[j],X[j+k]);k:=k shl %l.1195 úBE
1 end end;\newline %l.1196 úBE
inc(X[n-1],X[n] shr 16); X[n]:=(X[n] and \$FFFF)+c;\newline %l.1197 úBE
inc(X[n0-1],X[n0] shr 16); X[n0]:=(X[n0] and \$FFFF)+c;\newline %l.1198 úBE
for j:=n downto 2 do begin\newline %l.1199 úBE
\hspace*{1cm}inc(X[j-1],X[j] shr 16); X[j]:=X[j] and\hspace*{1ex}\$FFFF %l.1200 úBE
end;\newline %l.1201 úBE
X[1]:=X[1] and \$FFFF\newline %l.1202 úBE
end;\newline %l.1203 úBE
begin for i:=1 to n do X[i]:=0; c:=0; even:=true end.%l.1204 úBE

\vspace{1ex}%l.1206 úBE
\noindent{}\rm The corresponding mixed multiplicative generator is %l.1207 úBE
obtained by omitting or changing the lines containing {\tt c}. On %l.1208 úBE
a 100MHz Pentium computer this (not optimized) program produces %l.1209 úBE
19~563 random numbers per second whereas 20~938 mixed multiplicative %l.1210 úBE
random numbers can be produced. A loss of speed of about 6.6{\%} %l.1211 úBE
seems us worth the gain of better random numbers. Note that the %l.1212 úBE
number {\tt c} suffers an overflow every about $ 750~000$th random %l.1213 úBE
number. This does not affect randomness and it is not worth the %l.1214 úBE
effort to correct this flaw. %l.1215 úBE
\item{}We set $ P=2^{31}-1$, $ r=8$ which leads to $ P^r-1=2^{34%l.1216 ó,ò...
}\cdot 3^{2}\cdot 5\cdot 7\cdot 11\cdot 17\cdot 31\cdot 41%l.1217 $...
\cdot 151\cdot 331\cdot 733\cdot 1709\cdot 21529\cdot 368140581013%l.1218 $...
\cdot 708651694622727115232673724657$. Moreover we take $ a_{r-1%l.1219 $...
}=1$, $ a_{r-2}={\ldots}=a_{1}=0$, $ a_{0}=60~045$ yielding %l.1220 úBE
\begin{eqnarray}%l.1221 ú$$...
\label{58}\hspace{-.7cm}&&X_{0}=1, X_{-1}={\ldots}=X_{-7}=0%l.1222 ú$$...
\hspace{.6ex},\hspace{2ex}X_{k+1}=X_{k}+60~045X_{k-7}{\rm\hspace%l.1223 `...
{.38ex}mod\hspace{.38ex}}2^{31}-1\hspace{.6ex},\\ %l.1224 ú$$...
\hspace{-.7cm}&&\NX =P^{8}\!-\!1\approx 2^{248},\hspace{ 3pt} Q_{%l.1225 ó,ò...
1}={\ldots}=Q_{8}=\left( \!P^{8}\!\right) \!^{1.00202-1},%l.1226 ú$$...
\hspace{ 3pt} Q_{9}={\ldots}=Q_{16}=\left( \!P^{8}\!\right) \!^{%l.1227 ó,ò...
0.06368-1}.\nonumber %l.1228 ú$$...
\end{eqnarray} %l.1229 úBE
The following program gives on a 100MHz Pentium 74~473 random numbers %l.1230 úBE
({\tt X[k]}) per second.%l.1231 úBE

\tt unit random2;\newline %l.1233 úBE
interface\newline %l.1234 úBE
var X:array[0..7] of longint; k:integer;\newline %l.1235 úBE
procedure nextrandom;\newline %l.1236 úBE
implementation\newline %l.1237 úBE
const a0=60045;\newline %l.1238 úBE
var i:integer; x0,x1,x2:longint;\newline %l.1239 úBE
procedure nextrandom;\newline %l.1240 úBE
begin\newline %l.1241 úBE
x0:=X[(k+1) and 7];\newline %l.1242 úBE
x2:=(x0 and \$FFFF)*a0; x1:=(x0 shr 16)*a0+(x2 shr 16);\newline %l.1243 úBE
x2:=(x2 and \$FFFF)+(x1 shr 15)+((x1 and \$7FFF) shl 16);\newline %l.1244 úBE
if (x2 shr 31)=1 then x2:=(x2 xor \$80000000)+1;\newline %l.1245 úBE
inc(x2,X[k]);\newline %l.1246 úBE
while (x2 shr 31)=1 do x2:=(x2 xor \$80000000)+1;\newline %l.1247 úBE
k:=(k+1) and 7;\newline %l.1248 úBE
if x2=\$7FFFFFFF then X[k]:=0 else X[k]:=x2\newline %l.1249 úBE
end;\newline %l.1250 úBE
begin k:=0; X[0]:=1; for i:=1 to 7 do X[i]:=0 end.%l.1251 úBE
\end{enumerate}%l.1252

\section{Results and outlook} %l.1254
We have generalized the spectral test. As the new feature we analyze %l.1255
the sequence of random numbers (I in the figures) not only the distribution %l.1256
of $n$-tupels (II in the figures).%l.1257

We saw that the mixed multiplicative generator did not pass the test %l.1259
with an ideal result. We were able to construct an improved generator %l.1260
which has the recursion formula %l.1261
\begin{equation}%l.1262 ú$
\label{54}X_{0}=0\hspace{.6ex},\hspace{2ex}X_{k+1}=aX_{k}+c{\rm\hspace%l.1263 `...
{.38ex}int}\left( k/2\right) {\rm\hspace{.38ex}mod\hspace{.38ex}}2^d%l.1264 ú$
\hspace{.6ex}.%l.1265 ú$
\end{equation} %l.1266
For the choice of the parameters $a$, $c$, $d$ we made suggestions %l.1267
in Eq.\ (\ref{50}). This generator (or the multiply recursive generator %l.1268
given in Eq.\ (\ref{58})) seems us to be the best choice in quality %l.1269
and performance. An implementation of a generator of this type with %l.1270
modulus $ 2^d=2^{256}\approx 10^{77}$ was presented in Ex.\ 7.1. %l.1271
The calculation of random numbers is fast even though the modulus %l.1272
is that large. We think that for all practical purposes pseudo random %l.1273
numbers generated with this generator can not be distinguished from %l.1274
a true random sequence.%l.1275

We were able to analyze this and several other generators. The choice %l.1277
of parameters was discussed in Sec.\ \ref{param}.%l.1278

\vspace{1ex}%l.1280
\noindent{}For practical purposes there is essentially no need for %l.1281
further improvements. From a purely mathematical point of view however %l.1282
there are lots of open questions.%l.1283

Some further generators are discussed in \cite{Schnetz}. However %l.1285
there is still little known about multiplicative generators with %l.1286
prime number modulus and a non-primitive multiplier. In this case %l.1287
$ N|\hat{g}|^{2}(s_{0},s_{1})$ is given as zero of the polynomial %l.1288
\begin{equation}%l.1289 ú$
\label{57}P_{s_{0}}\left( Y\right) \equiv \prod _{s_{1}\in %l.1290 { ú$
{\ErgoBbb Z}_M}\left( Y-N|\hat{g}|^{2}\left( s_{0},s_{1}%l.1291 (,)...
\right) \right) \hspace{.6ex}.%l.1292 ú$
\end{equation} %l.1293
For multiplicative generators $ P_{s_{0}}(Y)=Y^M-MNY^{M-1}+%l.1294 $
{\ldots}$. Numerical calculations show that $ P_{s_{0}}$ has integer %l.1295
coefficients. We were not able to prove this for $ s_{0}\neq 0$ %l.1296
nor to analytically determine the coefficients for non-trivial examples.%l.1297

Further on, the Fourier analysis of generators involving polynomials %l.1299
may lead to interesting results. Here exist some connections to %l.1300
the theory of Gau{\ss} sums.%l.1301

Finally we would be interested in multiply recursive generators. %l.1303
Those generators are given by a matrix-valued multiplier. The simplest %l.1304
example with a prime number modulus was presented in Sec.\ \ref{multrec}. %l.1305
In this section we saw that multiply recursive generators are also %l.1306
the best candidates for being even more efficient than the generator %l.1307
given in (\ref{54}). In this connection multiply recursive generators %l.1308
with power of two modulus may be of special interest.%l.1309

\section*{Aknowledgement} %l.1311
I am grateful to Manfred H{\"u}ck who motivated me to this work by %l.1312
showing me some figures of random number generators.%l.1313

\section*{Figures} %l.1315
Some graphs of random number generators are presented to give a visual %l.1316
impression of what the generator looks like. There are two possibilities %l.1317
to draw a two-dimensional plot: first (I), to plot the $k$-th random %l.1318
number $X$$_{k}$ over $k$ and second (II), to plot $X$$_{k+1}$ over %l.1319
$X$$_{k}$ presenting the pair correlation. The third part of the %l.1320
figures give the absolute of the Fourier transform of I. $ |%l.1321 $
\hat{g}|^{2}(s_{0},s_{1})$ is a measure for the correlations along %l.1322
a line perpendicular to $ (s_{0},s_{1})$ in I (cf.\ Eq.\ (\ref{55})). %l.1323
For ideal generators $ |\hat{g}|^{2}$ should be $ \le 1$ and Figs.\ %l.1324
I and II should look like first rain drops on a dry road.\vspace{1cm}%l.1325

\noindent{} %l.1327
\unitlength0.027mm %l.1328
\begin{tabular}{crcrcr}\multicolumn{5}{l}{Fig.\ 1: $ X_{0}=1$, $ X%l.1329 $...
_{k+1}=195X_{k}{\rm\hspace{.38ex}mod\hspace{.38ex}}1009$}&\\ %l.1330 úBT
\raisebox{ 2.43cm}{\hspace{-3mm}$X$$_{k}$}&\hspace{-3ex}\input{a195a0.100%l.1331 úBT
}&\hspace{-1.5ex}\raisebox{ 2.43cm}{$X$$_{k+1}$}&\hspace{-3ex}\input%l.1332 úBT
{b195a0.100}&\hspace{-5ex}\raisebox{ 2.43cm}{$s$$_{1}$}\hspace{-2.7ex} %l.1333 úBT
0&\hspace{-6.4ex}%l.1334 úBT...
\begin{tabular}[b]{ccccc}$\frac{1}{1008}$&\hspace{-1ex}$\frac{1009}{1008}$%l.1335 úBT...
&\hspace{-1ex}$\frac{1009}{1008}$&\hspace{-1ex}$\frac{1009}{1008}$%l.1336 úBT...
&\hspace{-1ex}$\frac{1009}{1008}$\\ %l.1337 úBT...
$\frac{1}{1008}$&\hspace{-1ex}$\frac{1009}{1008}$&\hspace{-1ex}$\frac{1009}{1008}$%l.1338 úBT...
&\hspace{-1ex}$\frac{1009}{1008}$&\hspace{-1ex}$\frac{1009}{1008}$\\ %l.1339 úBT...
$\frac{1}{1008}$&\hspace{-1ex}$\frac{1009}{1008}$&\hspace{-1ex}$\frac{1009}{1008}$%l.1340 úBT...
&\hspace{-1ex}$\frac{1009}{1008}$&\hspace{-1ex}$\frac{1009}{1008}$\\ %l.1341 úBT...
$\frac{1}{1008}$&\hspace{-1ex}$\frac{1009}{1008}$&\hspace{-1ex}$\frac{1009}{1008}$%l.1342 úBT...
&\hspace{-1ex}$\frac{1009}{1008}$&\hspace{-1ex}$\frac{1009}{1008}$\\ %l.1343 úBT...
1008&0\hspace*{ 1ex}&0\hspace*{ 1ex}&0\hspace*{ 1ex}&0\hspace*{ 1ex}%l.1344 úBT...
\end{tabular}\\ %l.1345 úBT
\hspace{ 5ex} I&$k$&\hspace{ 3.5ex} II&$X$$_{k}$&$ |\hat{g}|^{2}%l.1346 $...
$\hspace{ 2.5mm}0&$s$$_{0}$\hspace*{ 2.5mm}%l.1347 úBT
\end{tabular}\vspace{1cm}%l.1348

\noindent{}%l.1350 úBT
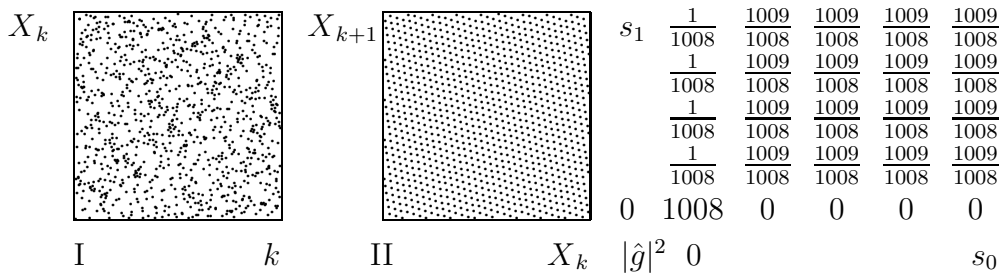
\begin{tabular}{crcrcr}\multicolumn{5}{l}{Fig.\ 2: $ X_{0}=0$, $ X%l.1351 $...
_{k+1}=37X_{k}+1{\rm\hspace{.38ex}mod\hspace{.38ex}}1024$}&\\ %l.1352 úBT
\raisebox{ 2.5cm}{\hspace{-3mm}$X$$_{k}$}&\hspace{-3ex}\input{a37b1.10%l.1353 úBT
}&\hspace{-1.5ex}\raisebox{ 2.5cm}{$X$$_{k+1}$}&\hspace{-3ex}\input%l.1354 úBT
{b37b1.10}&\hspace{-3ex}\raisebox{ 2.5cm}{$s$$_{1}$}\hspace{-2.7ex} %l.1355 úBT
0&\hspace{-5ex}%l.1356 úBT...
\begin{tabular}[b]{ccccc}0&0&0&0&16\\ %l.1357 úBT...
0&4&0&0&0\\ %l.1358 úBT...
0&0&8&0&0\\ %l.1359 úBT...
0&4&0&0&0\\ %l.1360 úBT...
2$^{10}$&0&0&0&0%l.1361 úBT...
\end{tabular}\\ %l.1362 úBT
\hspace{ 5ex} I&$k$&\hspace{ 3.5ex} II&$X$$_{k}$&$ |\hat{g}|^{2}%l.1363 $...
$\hspace{ 1mm}0&$s$$_{0}$\hspace*{ 1.5mm}%l.1364 úBT
\end{tabular}\vspace{1cm}%l.1365

\noindent{}%l.1367 úBT
\begin{tabular}{crcrcr}\multicolumn{5}{l}{Fig.\ 3: $ X_{0}=0$, $ X%l.1368 $...
_{k+1}=41X_{k}+3{\rm\hspace{.38ex}mod\hspace{.38ex}}1024$}&\\ %l.1369 úBT
\raisebox{ 2.5cm}{\hspace{-3mm}$X$$_{k}$}&\hspace{-3ex}\input{a41b3.10%l.1370 úBT
}&\hspace{-1.5ex}\raisebox{ 2.5cm}{$X$$_{k+1}$}&\hspace{-3ex}\input%l.1371 úBT
{b41b3.10}&\hspace{-3ex}\raisebox{ 2.5cm}{$s$$_{1}$}\hspace{-2.7ex} %l.1372 úBT
0&\hspace{-5ex}%l.1373 úBT...
\begin{tabular}[b]{ccccc}0&0&0&0&32\\ %l.1374 úBT...
0&8&0&0&0\\ %l.1375 úBT...
0&0&16&0&0\\ %l.1376 úBT...
0&8&0&0&0\\ %l.1377 úBT...
2$^{10}$&0&0&0&0%l.1378 úBT...
\end{tabular}\\ %l.1379 úBT
\hspace{ 5ex} I&$k$&\hspace{ 3.5ex} II&$X$$_{k}$&$ |\hat{g}|^{2}%l.1380 $...
$\hspace{ 1mm}0&$s$$_{0}$\hspace*{ 1.5mm}%l.1381 úBT
\end{tabular}\vspace{1cm}%l.1382

\noindent{}%l.1384 úBT
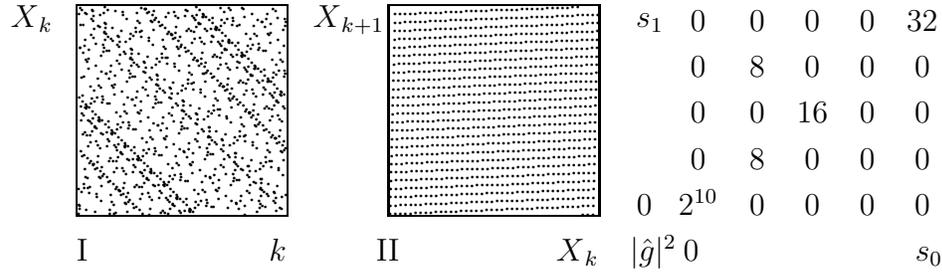
\begin{tabular}{crcrcr}\multicolumn{5}{l}{Fig.\ 4: $ X_{0}=0$, $ X%l.1385 $...
_{k+1}=41X_{k}+1{\rm\hspace{.38ex}mod\hspace{.38ex}}1024$}&\\ %l.1386 úBT
\raisebox{ 2.5cm}{\hspace{-3mm}$X$$_{k}$}&\hspace{-3ex}\input{a41b1.10%l.1387 úBT
}&\hspace{-1.5ex}\raisebox{ 2.5cm}{$X$$_{k+1}$}&\hspace{-3ex}\input%l.1388 úBT
{b41b1.10}&\hspace{-3ex}\raisebox{ 2.5cm}{$s$$_{1}$}\hspace{-2.7ex} %l.1389 úBT
0&\hspace{-5ex}%l.1390 úBT...
\begin{tabular}[b]{ccccc}0&0&0&0&0\\ %l.1391 úBT...
0&8&0&0&0\\ %l.1392 úBT...
0&0&0&0&0\\ %l.1393 úBT...
0&0&0&8&0\\ %l.1394 úBT...
2$^{10}$&0&0&0&0%l.1395 úBT...
\end{tabular}\\ %l.1396 úBT
\hspace{ 5ex} I&$k$&\hspace{ 3.5ex} II&$X$$_{k}$&$ |\hat{g}|^{2}%l.1397 $...
$\hspace{ 1mm}0&$s$$_{0}$\hspace{ 1.5mm}%l.1398 úBT
\end{tabular}\vspace{1cm}%l.1399

\noindent{}%l.1401 úBT
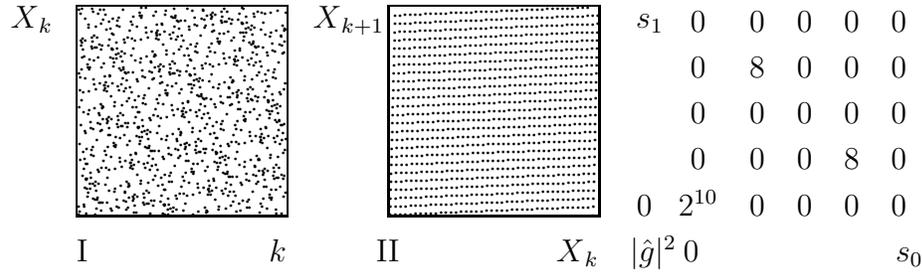
\begin{tabular}{crcrcr}\multicolumn{5}{l}{Fig.\ 5: $ X_{0}=0$, $ X%l.1402 $...
_{k+1}=21X_{k}+1{\rm\hspace{.38ex}mod\hspace{.38ex}}1000$}&\\ %l.1403 úBT
\raisebox{ 2.4cm}{\hspace{-3mm}$X$$_{k}$}&\hspace{-3ex}\input{a21b1.100%l.1404 úBT
}&\hspace{-1.5ex}\raisebox{ 2.4cm}{$X$$_{k+1}$}&\hspace{-3ex}\input%l.1405 úBT
{b21b1.100}&\hspace{-4ex}\raisebox{ 2.4cm}{$s$$_{1}$ 0}\hspace{-2.7ex}%l.1406 úBT
$ -4$&\hspace{-6.3ex}%l.1407 úBT...
\begin{tabular}[b]{ccccc}1000&0&0&0&0\\ %l.1408 úBT...
0&0&0&0&0\\ %l.1409 úBT...
0&0&40&0&0\\ %l.1410 úBT...
0&0&0&0&0\\ %l.1411 úBT...
0&0&0&0&40%l.1412 úBT...
\end{tabular}\\ %l.1413 úBT
\hspace{ 5ex} I&$k$&\hspace{ 3.5ex} II&$X$$_{k}$&$ \hspace{ 1.5mm%l.1414 $...
}|\hat{g}|^{2}$\hspace{ 4.5mm}0&$s$$_{0}$\hspace*{ 1.5mm}%l.1415 úBT
\end{tabular}\vspace{1cm}%l.1416

\noindent{}%l.1418 úBT
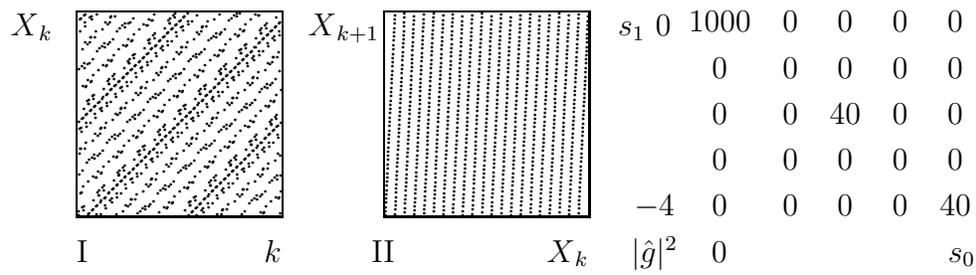
\begin{tabular}{crcrcr}\multicolumn{5}{l}{Fig.\ 6: $ X_{0}=0$, $ X%l.1419 $...
_{k+1}=(37+1024)X_{k}+1{\rm\hspace{.38ex}mod\hspace{.38ex}}1024^{%l.1420 ó,ò...
2}$\vspace{ 2mm}}&\\ %l.1421 úBT
\raisebox{ 2.5cm}{\hspace{-3mm}$X$$_{k}$}&\hspace{-3ex}\input{a1061e1.10%l.1422 úBT
}&\hspace{-1.5ex}\raisebox{ 2.5cm}{$X$$_{k+1}$}&\hspace{-3ex}\input%l.1423 úBT
{b1061e1.10}&\hspace{-5.5ex} &\hspace{-4ex}%l.1424 úBT...
\begin{tabular}[b]{c}cf. Fig.\ 2\\ %l.1425 úBT...
with $ \NX =10^{20}$%l.1426 úBT...
\end{tabular}\\ %l.1427 úBT
\hspace{ 5ex} I&$k$&\hspace{ 3.5ex} II&$X$$_{k}$&&%l.1428 úBT
\end{tabular}\vspace{1cm}%l.1429

\noindent{}%l.1431 úBT
\begin{tabular}{crcrcr}\multicolumn{5}{l}{Fig.\ 7: $ X_{0}=0$, $ X%l.1432 $...
_{k+1}=37X_{k}+129{\rm\hspace{.38ex}int}(k/2){\rm\hspace%l.1433 `...
{.38ex}mod\hspace{.38ex}} 1024$}&\\ %l.1434 úBT
\raisebox{ 2.5cm}{\hspace{-3mm}$X$$_{k}$}&\hspace{-3ex}\input{a37d129.10%l.1435 úBT
}&\hspace{-1.5ex}\raisebox{ 2.5cm}{$X$$_{k+1}$}&\hspace{-3ex}\input%l.1436 úBT
{b37d129.10}&\hspace{-3ex}\raisebox{ 2.5cm}{$s$$_{1}$}\hspace{-2.7ex} %l.1437 úBT
0&\hspace{-5.5ex}%l.1438 úBT...
\begin{tabular}[b]{ccccc}4&0&0&0&4\\ %l.1439 úBT...
1&1&1&1&1\\ %l.1440 úBT...
2&0&2&0&2\\ %l.1441 úBT...
1&1&1&1&1\\ %l.1442 úBT...
2$^{11}$&0&0&0&0%l.1443 úBT...
\end{tabular}\\ %l.1444 úBT
\hspace{ 5ex} I&$k$&\hspace{ 3.5ex} II&$X$$_{k}$&$ |\hat{g}|^{2}%l.1445 $...
$\hspace{ 1.8mm}0&$s$$_{0}$\hspace*{ 1mm}%l.1446 úBT
\end{tabular}\vspace{1cm}%l.1447

\noindent{}%l.1449 úBT
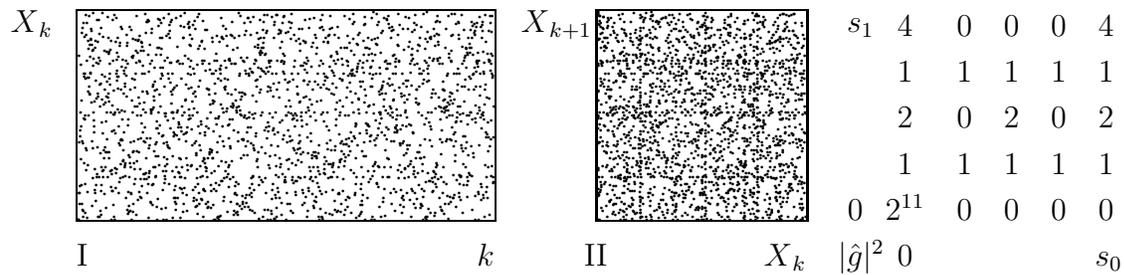
\begin{tabular}{crcrcr}\multicolumn{5}{l}{Fig.\ 8: $ X_{0}=1$, $ X%l.1450 $...
_{-1}=0$, $ X_{k+1}=X_{k}+7X_{k-1}{\rm\hspace{.38ex}mod\hspace{.38ex}}31%l.1451 $...
$}&\\ %l.1452 úBT
\raisebox{ 2.2cm}{\hspace{-3mm}$X$$_{k}$}&\hspace{-3ex}\input{a1f7.31%l.1453 úBT
}&\hspace{-1.5ex}\raisebox{ 2.2cm}{$X$$_{k+1}$}&\hspace{-3ex}\input%l.1454 úBT
{b1f7.31}&\hspace{-3ex}\raisebox{ 2.5cm}{$s$$_{1}$}\hspace{-2.7ex} %l.1455 úBT
0&\hspace{-7.5ex}%l.1456 úBT...
\begin{tabular}[b]{ccccc}$\frac{1}{960}$&\hspace{-1ex}$\frac{961}{960}$%l.1457 úBT...
&\hspace{-1ex}$\frac{961}{960}$&\hspace{-1ex}$\frac{961}{960}$&\hspace{-1ex}$\frac{961}{960}$\\ %l.1458 úBT...
$\frac{1}{960}$&\hspace{-1ex}$\frac{961}{960}$&\hspace{-1ex}$\frac{961}{960}$%l.1459 úBT...
&\hspace{-1ex}$\frac{961}{960}$&\hspace{-1ex}$\frac{961}{960}$\\ %l.1460 úBT...
$\frac{1}{960}$&\hspace{-1ex}$\frac{961}{960}$&\hspace{-1ex}$\frac{961}{960}$%l.1461 úBT...
&\hspace{-1ex}$\frac{961}{960}$&\hspace{-1ex}$\frac{961}{960}$\\ %l.1462 úBT...
$\frac{1}{960}$&\hspace{-1ex}$\frac{961}{960}$&\hspace{-1ex}$\frac{961}{960}$%l.1463 úBT...
&\hspace{-1ex}$\frac{961}{960}$&\hspace{-1ex}$\frac{961}{960}$\\ %l.1464 úBT...
960&0\hspace*{ 1ex}&0\hspace*{ 1ex}&0\hspace*{ 1ex}&0\hspace*{ 1ex}%l.1465 úBT...
\end{tabular}\\ %l.1466 úBT
\hspace{ 5ex} I&$k$&\hspace{ 3.5ex} II&$X$$_{k}$&$ |\hat{g}|^{2}%l.1467 $...
$\hspace{ 3.5mm}0&$s$$_{0}$\hspace*{ 1.5mm}%l.1468 úBT
\end{tabular}%l.1469

%l.1492
\end{document}